\documentstyle{mn}

\def\simlt{\mathrel{\rlap{\lower 3pt\hbox{$\sim$}}
        \raise 2.0pt\hbox{$<$}}}
\def\simgt{\mathrel{\rlap{\lower 3pt\hbox{$\sim$}}
        \raise 2.0pt\hbox{$>$}}}          
\def\whzsr{\,{\rm W\, Hz^{-1} sr^{-1}}}

\begin{document}
\title[The 2dF Galaxy Redshift Survey: The population of nearby radio 
galaxies at the 1 mJy level]
{The 2dF Galaxy Redshift Survey: The population of nearby radio galaxies at 
the 1 mJy level}
\author[Manuela Magliocchetti et al.]{
\parbox[t]{\textwidth}{
Manuela Magliocchetti$^{1}$,
Steve J.\ Maddox${^2}$, Carole A.\ Jackson${^3}$, 
Joss Bland-Hawthorn$^4$,
Terry Bridges$^4$, 
Russell Cannon$^4$, 
Shaun Cole$^5$, 
Matthew Colless$^3$, 
Chris Collins$^6$, 
Warrick Couch$^7$, 
Gavin Dalton$^9$,
Roberto de Propris$^7$,
Simon P.\ Driver$^8$, 
George Efstathiou$^{10}$, 
Richard S.\ Ellis$^{11}$, 
Carlos S.\ Frenk$^4$, 
Karl Glazebrook$^{12}$, 
Ofer Lahav$^{10}$, 
Ian Lewis$^4$, 
Stuart Lumsden$^{13}$, 
John A.\ Peacock$^{14}$,
Bruce A.\ Peterson$^3$, 
Will Sutherland$^{14}$,
Keith Taylor$^3$}
\vspace*{6pt} \\ 
$^1$SISSA, Via Beirut 4, 34100, Trieste, Italy \\
$^2$School of Physics and Astronomy, University of Nottingham,
Nottingham NG7 2RD, UK\\
$^3$Research School of Astronomy and Astrophysics,
 The Australian National University, Canberra, ACT 2611, Australia\\    
$^4$Anglo-Australian Observatory, P.O.\ Box 296, Epping, NSW 2121,
    Australia\\  
$^5$Department of Physics, University of Durham, South Road, 
    Durham DH1 3LE, UK \\ 
$^6$Astrophysics Research Institute, Liverpool John Moores University,  
    Twelve Quays House, Birkenhead, L14 1LD, UK \\
$^7$Department of Astrophysics, University of New South Wales, Sydney, 
    NSW 2052, Australia \\
$^8$School of Physics and Astronomy, University of St Andrews, 
    North Haugh, St Andrews, Fife, KY6 9SS, UK \\
$^9$Department of Physics, University of Oxford, Keble Road, 
    Oxford OX1 3RH, UK \\
$^{10}$Institute of Astronomy, University of Cambridge, Madingley Road,
    Cambridge CB3 0HA, UK \\
$^{11}$Department of Astronomy, California Institute of Technology, 
    Pasadena, CA 91125, USA \\
$^{12}$Department of Physics \& Astronomy, Johns Hopkins University,
       Baltimore, MD 21218-2686, USA \\
$^{13}$Department of Physics, University of Leeds, Woodhouse Lane,
       Leeds, LS2 9JT, UK \\
$^{14}$Institute for Astronomy, University of Edinburgh, Royal Observatory, 
       Blackford Hill, Edinburgh EH9 3HJ, UK \\
}
\maketitle\begin{abstract}
We use redshift determinations and spectral analysis of galaxies in the 
2dF Galaxy Redshift Survey to study the properties of local radio sources 
with $S\ge 1$~mJy. 557 objects (hereafter called  the
spectroscopic sample) drawn from the FIRST survey, corresponding to 
$2.3$ per cent of the total radio sample, are found in the 2dFGRS 
catalogue within the area $9^h 48^m \simlt {\rm RA}({\rm 2000}) 
\simlt 14^h 32^m$ and $-2.77^\circ \simlt {\rm dec}({\rm 2000}) \simlt 2.25^\circ$, 
down to a magnitude limit $b_J= 19.45$. 
The excellent quality of 2dF spectra allows us to divide 
these sources into classes, according to their optical spectra. 

Absorption line systems make up 63 per cent of the spectroscopic
sample. These may or may not show emission lines due to AGN activity and
correspond to ``classical'' radio galaxies belonging mainly to the FRI
class. They are characterized by relatively high radio-to-optical
ratios, red colours, and high radio luminosities ($10^{21}\simlt 
P_{1.4{\rm GHz}}/ \whzsr \simlt 10^{24}$).
Actively star-forming galaxies contribute about 32 per cent of the
sample. These objects are mainly found at low redshifts ($z \simlt 0.1$) and
show low radio-to-optical ratios, blue colours and low radio
luminosities.  We also found 18 Seyfert 2 galaxies (3 per cent) and 4
Seyfert 1's (1 per cent).
                           

Analysis of the local radio luminosity function shows that radio galaxies 
are well described by models that assume pure luminosity evolution, 
at least down to radio powers $P_{1.4{\rm GHz}}\simlt 10^{20.5} \whzsr$. 
Late-type galaxies, whose relative contribution to the RLF
is found to be lower than was predicted by previous works, present a luminosity function 
which is comparable with the IRAS galaxy LF.
This class of sources therefore plausibly constitutes the radio counterpart 
of the dusty spirals and starbursts that dominate the counts at 60~$\mu$m. 
\end{abstract}

\section{Introduction}

During the last twenty years, several attempts have been made to model the 
space density and evolution of radio sources. These attempts have mainly 
followed two well defined tracks, where the first class of models 
(Orr \& Brown 1982; Padovani
\& Urry 1992; Maraschi \& Rovetti 1994, Wall \& Jackson 1997 just to mention 
a few) base their predictions on the unification paradigm which
originates from the `relativistic jet' model of Blandford \& Rees
(1978), while the second one (Wall, Pearson \& Longair 1980; Danese et
al. 1987; Dunlop \& Peacock 1990; Condon 1984; Rowan-Robinson et
al. 1993) relies on the evolutionary properties of the galaxy hosting
the radio source.

One of the main limitations affecting both classes of models
(especially those from the unification paradigm) is the fact that they
were mainly based on datasets including very bright sources
($S_{1.4\rm GHz} \simeq 1$~Jy). As a consequence, the low-power tail
of the AGN radio luminosity function is poorly defined, and their
predictions at faint flux densities diverge. Some of the models
belonging to the second class (e.g. Danese et al. 1987; Condon 1984;
Rowan-Robinson et al. 1993) have pushed their analysis down to much
lower flux densities ($S\sim 0.1$~mJy), and assume the contribution to
the radio population at $S\simlt 10$~mJy to be principally given by a
new class of objects which greatly differs from the radio AGN which
dominate at higher fluxes. The limited statistical samples that are
available at such faint flux densities, mean that the nature of this
population has remained an issue of debate. For instance, Condon
(1984) suggests a population of strongly-evolving normal spiral
galaxies, while others (Windhorst et al. 1985; Danese et al. 1987;
Rowan-Robinson et al. 1993) claim the presence of an actively
star-forming galaxy population.

More recently, great efforts have been made to determine the
photometric and spectroscopic properties of radio sources at mJy
levels and fainter (Gruppioni et al. 1999; Georgakakis et al. 1999;
Magliocchetti et al. 2000; Masci et al. 2001). These studies have
proven extremely useful in characterizing the populations of radio
sources but, once again, the limited statistical samples over small
areas of the sky leave significant uncertainties such as the relative
contribution of actively star-forming galaxies to the total mJy
counts.

The advent of large area radio surveys which probe the radio sky down
to mJy levels (e.g. FIRST, Becker, Helfand \& White 1995; NVSS, Condon et al.
1998; SUMSS, Bock et al. 1999) overcomes these small sample
limitations which have affected studies of faint radio sources up to
now. This can be achieved with multi-wavelength follow-ups of large
samples of sources drawn from such wide-area surveys. In particular,
the acquisition of optical spectra enables to derive spectral types,
luminosity functions and redshift distributions for the different
radio-populations.

The first results in this direction are given by Sadler et al. (1999)
who use spectra from the 2dF Galaxy Redshift Survey (Maddox 1998, Colless et 
al. 2001) to
investigate the nature of 127 radio sources drawn from the NVSS survey
down to $\sim 2.5$~mJy. More recently this analysis has been extended
to include 912 candidate optical counterparts to NVSS sources (Sadler
et al. 2002).  The present paper follows a similar approach, and
presents results for a subsample of 557 sources taken from the FIRST
radio catalogue.  The FIRST survey includes radio sources that are up to 3
times fainter than the NVSS, and the accuracy of the FIRST radio
positions is much better than the NVSS, which means that the likelihood
of correct identifications is significantly improved. On the other
hand, the NVSS overlaps a much larger area of the 2dF survey, giving a
larger sample of objects to study.  For both samples, the 2dF survey
spectra provide information on the composition of the radio population
associated with optical counterparts brighter than $b_J=19.45$.
Redshift determinations further allow us to estimate quantities such
as luminosity functions and redshift distributions for the different
classes of sources contributing to the local population (the 2dF
Galaxy Redshift Survey samples galaxies up to $z\simeq 0.3$; Colless
et al. 2001).

The layout of the paper is as follows: Section 2 briefly introduces the 
surveys which provided the data (Section 2.1 and 2.2) and describes the 
procedure we adopted to obtain the spectroscopic counterparts of a subsample 
of FIRST radio sources (Section 2.3). Section 3 illustrates the optical and 
spectroscopic properties of the objects in the sample, while Section 4 deals 
with their radio properties and present the results for the local radio 
luminosity function, obtained by comparing our data with model predictions for 
different classes of sources. Finally, Section 5 is devoted to the analysis 
of the observed redshift distributions and Section 6 summarizes our
conclusions. Throughout the paper we will assume $\Omega_0=1$ and 
$h_0=0.5$, where $H_0=h_0\times 100\,{\rm km\, s^{-1}\, Mpc^{-1}}$.   
   
\section{The Datasets}
\subsection{The 2dFGRS}
The 2dF Galaxy Redshift Survey (2dFGRS: Maddox 1998; Colless et al. 2001) 
is a large-scale survey aimed at obtaining spectra for 250\,000 galaxies 
to an extinction-corrected limit for completeness of $b_J=19.45$ over 
an area of 2151 square degrees. 
The survey geometry consists of two broad declination strips, a larger one 
in the SGP covering the area $3^h 30^m\simlt {\rm RA}({\rm 2000})\simlt 
21^h40^m$, $-37.5^\circ \simlt {\rm dec}({\rm 2000})\simlt -22.5^\circ$ and a smaller one 
set in the NGP with $9^h 50^m\simlt {\rm RA}({\rm 2000})\simlt 14^h50^m$, 
$2.5^\circ \simlt{\rm dec}({\rm 2000}) \simlt -7.5^\circ$, plus 100 random 2-degree 
fields spread uniformly over the 7000 square degrees of the APM catalogue in 
the southern Galactic hemisphere.

The input catalogue for the survey is a revised version of the APM galaxy 
catalogue (Maddox et al. 1990a, 1990b, 1996) which includes over 5 
million galaxies down to $b_J=20.5$ in both north and south Galactic 
hemispheres over a region of almost 10$^4$ square degrees.
The astrometric rms error for galaxies with $17\le b_J\le 19.5$ is 0.25~arcsec,
while the photometry of the catalogue has a precision of about 0.15~mag for 
galaxies within the same magnitude limits. The mean surface brightness 
isophotal detection limit of the APM catalogue is $b_J=25\,\rm mag\, arcsec^{-2}$.

Redshifts for all the sources brighter than $b_J=19.45$ are determined
in two independent ways, via both cross-correlation of the spectra
with specified absorption-line templates (Colless et al. 2001) and by
emission-line fitting. These automatic redshift estimates have then
been confirmed by visual inspection of each spectrum, and the more
reliable of the two results chosen as the final redshift. A quality
flag was assigned to each redshift: $Q=3$, $Q=4$ and $Q=5$ correspond to reliable
redshift determination, $Q=2$ means a probable redshift and $Q=1$
indicates no redshift measurement.  The success rate in redshift
acquisition for the surveyed galaxies (determined by the inclusion in
the 2dF sample of only those objects with quality flags $Q=3$ to $Q=5$)
is estimated about 95 per cent (Folkes et al. 1999).  The median
redshift of the galaxies is 0.11 and the great majority have
$z<0.3$. In this work we use the version of the catalogue derived from
the November 1997 to January 2001 observations of the north galactic cap 
equatorial area, which includes 35,347 galaxies.

\subsection{The Radio Data}

The input catalogue for the radio data has been obtained by matching together 
sources in the FIRST and APM catalogues as extensively described in 
Magliocchetti \& Maddox (2002).

Briefly, the original radio catalogue comes from the FIRST (Faint Images of 
the Radio Sky at Twenty centimetres) survey (Becker et al. 1995).
The latest release (5 July 2000) of the catalogue covers about 7988 square 
degrees of the sky in the north Galactic cap and equatorial zones, including 
most of the area $7^h20^m \simlt {\rm RA}({\rm 2000}) \simlt 
17^h20^m$, $22.2^\circ \simlt {\rm dec}({\rm 2000}) \simlt 57.5^\circ$ and $21^h20^m 
\simlt{\rm RA}({\rm 2000}) \simlt 3^h20^m$, $-2.8^\circ \simlt {\rm dec}({\rm 2000}) 
\simlt 2.2^\circ$, and comprises approximately 722,354 sources down to a flux 
limit $S_{1.4 {\rm GHz}}\simeq 0.8$~mJy, with a 5$\sigma$ source detection 
limit of $\sim 1$~mJy. The catalogue has been estimated to be 95 per cent 
complete at 2~mJy and 80 per cent complete at 1~mJy (Becker et al.~1995).   

Optical counterparts for a sub-sample of FIRST radio sources have then been 
obtained by matching together objects included in the radio catalogue with 
objects coming from the APM survey. Since the APM and FIRST surveys only 
overlap in a relatively small region of the 
sky  on the equatorial plane between $9^h 48^m \simlt {\rm RA}({\rm 2000}) 
\simlt 14^h 32^m$ and $-2.77^\circ \simlt {\rm dec}({\rm 2000}) \simlt 2.25^\circ$, 
the search for optical counterparts was restricted to this area.

\begin{figure}
\vspace{8cm}  
\includegraphics{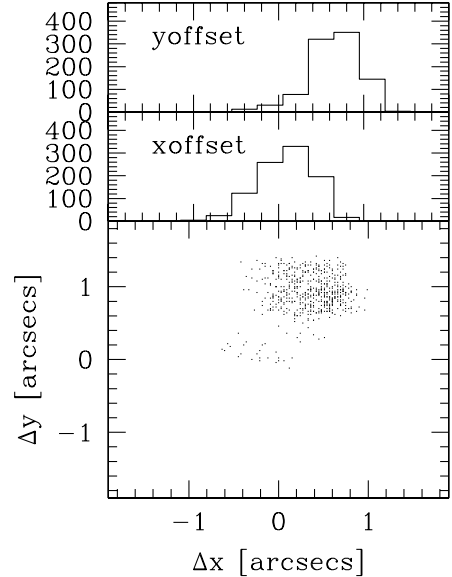}
\caption{Lower panel: distribution of the residuals $\Delta x=x_{\rm OPT}-
x_{\rm SPEC}$, $\Delta y=y_{\rm OPT}-y_{\rm SPEC}$ between optical and 
spectroscopic positions.
Middle and top panels: number of matches as a function of $\Delta x$ and
$\Delta y$ offsets.
\label{fig:allres}}   
\end{figure}

Out of approximately 24,000 radio sources with $S\ge 1$~mJy in the
considered area, Magliocchetti \& Maddox (2002) find 4075
identifications -- corresponding to 16.7 per cent of the original
sample -- in the APM catalogue for $b_J\le 22$ and for a matching
radius of 2 arcsec. This last value was chosen after a careful
analysis of the distribution of the residuals $\Delta x=x_{\rm
RADIO}-x_{\rm OPTICAL}$, $\Delta y=y_{\rm RADIO}-y_{\rm OPTICAL}$
between the positions of radio and optical sources. The rms value of
the $\Delta x$--$\Delta y$ distribution is found to be 0.7~arcsec,
which is consistent with the uncertainty obtained by summing the
positional uncertainties of the FIRST ($\sim 0.5$~arcsec) and APM
($\sim 0.5$~arcsec) surveys in quadrature and taking into account a
small distortion between the radio and optical reference frames.  So a
2 arcsec match radius is equivalent to about $2.5\sigma$, which should
include $\sim 97\% $ of the true identifications.  Furthermore, using
this match radius limits the number of random coincidences to a
negligible $\sim 5$ per cent (225 sources out of 4075). To ensure
uniform completeness over the optical survey, we restricted the
analysis to objects with $b_J\le 21.5$, leading to 3176
identifications.

Since the adopted version of the APM data includes magnitude
measurements only in the $b_j$ band, R magnitudes were then assigned
to the optical identifications by making use of the $b_J$-R colour
estimates given in APMCAT (found online at {\it
http://www.ast.cam.ac.uk/ $^\sim$apmcat} Irwin et al.). This version of
the APM data has been processed to provide accurate stellar
magnitudes, and so the galaxy magnitudes are not so reliable, but it
does include both R and $b_j$ UKST plates. Since the data are not
designed for galaxy work, these colours have a rather large
uncertainty, but nevertheless provide interesting information.

In the magnitude range $16\le b_J\le 20.5$ image profiles were used to
separate galaxies from stars as described by Maddox et al. (1990a). For
$b_J\ge 20.5$ galaxy profiles do not greatly differ from stellar
profiles and the classification becomes unreliable.  So, for $16\le
b_J\le 20.5$ we divide the radio sources into stellar-like objects,
expected to be mostly high redshift QSO (535 objects) and radio
galaxies (1494 objects), expected to lie within  $z \simeq 0.3$. The
radio-to-optical ratios and colours of the stellar identifications are
consistent with them being mostly high redshift QSOs.

\subsection{The Spectroscopic Sample}

In order to obtain redshift measurements and spectral features for
those radio sources with an optical counterpart (hereafter called
photometric sample) as described in Section 2.2, we looked for objects
in the 2dF catalogue with positions which differed by less than 2
arcsec from positions of sources in the photometric sample. The
choice of this value for the matching radius is based on the 2 arcsec
diameter of each 2dF fibre. Indeed we found all the matched objects to
have positional offsets of maximum 1 arcsec, with a rms of $\sim 0.5$
arcsec.  
Fig. 1 shows the distribution of the residuals $\Delta
x=x_{\rm OPT}- x_{\rm SPEC}$, $\Delta y=y_{\rm OPT}-y_{\rm SPEC}$
between the positions of sources in the photometric and 2dF
catalogues. The bottom panel represents the $\Delta x$-$\Delta y$
distribution, while the middle and top panels show the number of
matches respectively as a function of $\Delta x$ and $\Delta y$
offsets. As already discussed, all the offsets lie within $\Delta
x\simeq
\Delta y\sim 1$, with $\sim 0.5$~arcsec rms. The shift of the median offset 
of the distribution from the zero values, more prominent along the 
$\Delta y$ axis, is  due to adjustments in the astrometry for 
galaxies in the 2dF survey relative to those included in the earlier version 
of the APM catalogue from which the photometric sample was drawn.   

\begin{figure*}
\vspace{8cm}  
\includegraphics{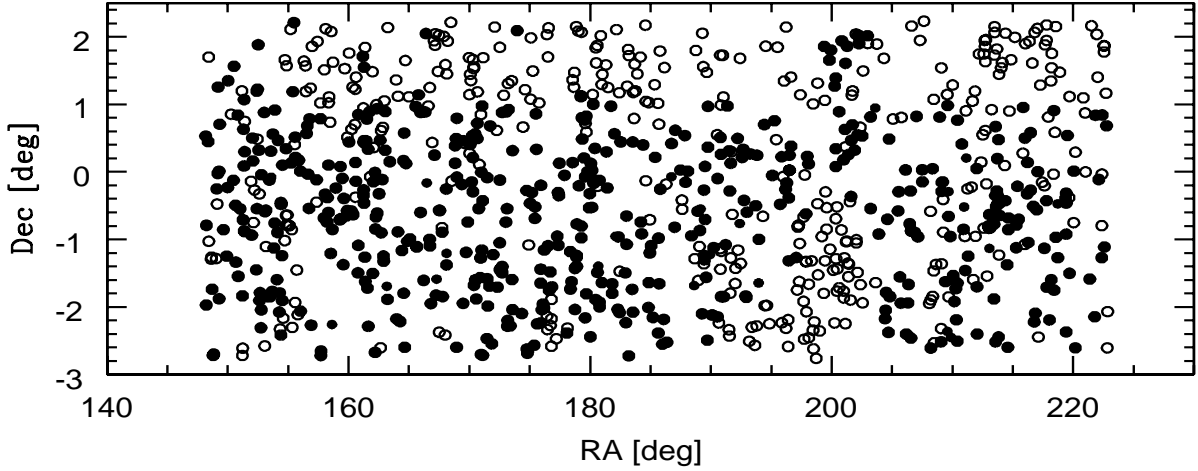}
\caption{Projected distribution of radio sources with optical counterparts 
in the APM catalogue with $ b_J\le 19.45$. 
Filled circles identify objects with spectral identifications, while empty 
ones are for those not yet included in the 2dFGRS.  
\label{fig:fields}}   
\end{figure*} 

Out of 971 input sources from the photometric catalogue with $b_J\le
19.45$ and $S\ge 1$~mJy, we were then able to obtain redshifts for 557
objects, 53 per cent of the original sample. Fig. 2 shows the
projected distribution of radio sources in the photometric sample with
$b_J\le 19.45$; filled circles identify those objects for which
redshift estimates and spectral classifications were available. We
remark here that the reason for such an apparent incompleteness in the
redshift determination is mainly due to incomplete sky coverage, as
there are areas which have not yet been observed by the 2dFGRS.  Less
that 40 objects from the input sample with $b_J\le 19.45$ could in
fact have been missed due to the completeness level of the redshift
survey ($\sim 95$ per cent; Colless et al. 2001).

To test for both radio-flux and magnitude biases in the determination
of the photometric and spectroscopic samples, we show the number of
radio sources respectively per flux and magnitude unit in Figs 3
and 4. In Fig. 3, the dotted line has been obtained from the
original FIRST catalogue for all those sources enclosed in the
equatorial region defined in Section 2, while the dashed line
illustrates the case for the photometric sample ($b_J\le 20.5$) and
the solid line represents the spectroscopic sample. In Fig. 4, the
dashed and solid lines are for the photometric and spectroscopic
samples respectively.  In each Figure the lower panel shows the ratio
of number counts in each sample. 
The ratio of spectroscopic to
photometric sources remains constant with magnitude up to $b_J\sim
19-19.5$, but there is a weak trend for there to be more
identifications at faint radio fluxes. 
It is not entirely clear what causes
this trend; one possibility could be the fact that fainter optical objects 
are in general associated with brighter radio sources 
(e.g. high-z ellipticals -- see later in Section 3 and 4). 
Whatever the reason, we can confidently exclude the presence of any strong 
bias in the
radio-flux or magnitude distribution, first of all for the optical
counterparts of radio sources (as already discussed in Magliocchetti
\& Maddox 2002) and also for the subset for which we have their
redshifts. As a final remark, note that Fig. 4 shows a tendency
for the number of sources both in the photometric and spectroscopic
samples to flatten for magnitudes $b_J\simgt 17.5$. As will be
discussed in Section 4, this behaviour is due to the absence of
optically faint star-forming galaxies in the FIRST survey.

Before carrying on with our analysis, it is worthwhile mentioning an important issue 
which needs to be taken into account when dealing with objects coming from the FIRST 
survey related to flux measurements. The high resolution of the survey in fact 
implies that some of the flux coming from extended sources could be either resolved 
out or split into two or more components, leading to a systematic underestimate of 
the real flux densities of such sources (Condon et al. 1998). This effect has been 
partially corrected for by using the method developed by Magliocchetti et al. (1998) 
to combine multicomponent sources. However, despite this correction, it turns out 
that fluxes as derived from the FIRST survey were on average still lower than those 
measured by the NVSS survey (which, having a lower resolution, should not be 
affected by this effect) by a factor of about 1~mJy in the case of sources brighter 
than 3~mJy, and about 30 per cent of the measured flux for sources with 
$1\,{\rm mJy}\le S_{1.4 {\rm GHz}}\simlt$3~mJy (Jackson et al. 2002). 
We therefore corrected the estimated flux densities by making use of the above 
quantities and, while for consistency reasons throughout the paper we will keep 
using the flux densities as measured by the FIRST survey, in the calculation of 
high-precision quantities such as the radio luminosity function (see Section 4), we 
will rely on these corrected values. It is nevertheless worth stressing that such 
corrections do not affect any of the results presented in this paper (including the 
luminosity functions) by more than a negligible factor, generally much lower than 
other uncertainties associated with the various measurements.

\begin{figure}
\vspace{8cm}  
\includegraphics{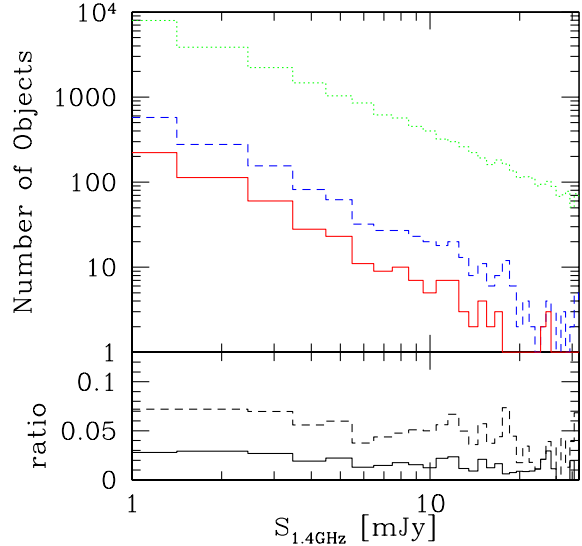}
\caption{Number of radio sources per flux interval in the considered area. 
The dotted line represents the original radio sample, the dashed line those 
sources with optical counterparts brighter than $b_J\le 22$ 
(photometric sample) while the solid line is for sources with spectral 
determinations from the 2dFGRS (spectroscopic sample). The lower panel 
represents the ratio between the number of sources respectively in the 
spectroscopic and radio sample (solid line) and optical and radio sample 
(dashed line) as a function of radio flux.
\label{fig:histfluxes}}   
\end{figure} 

\begin{figure}
\vspace{8cm}  
\includegraphics{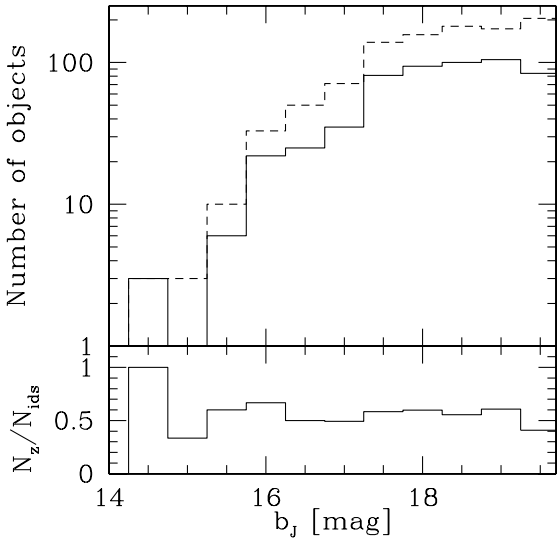}
\caption{Number of radio sources per magnitude interval over the considered 
area. The dashed line represents sources from the photometric sample while the 
solid line is for sources with spectral 
determinations from the 2dFGRS (spectroscopic sample). The lower panel 
represents the ratio between the number of sources respectively in the 
spectroscopic and photometric sample as a function of magnitude. 
\label{fig:histmags}}   
\end{figure}

\section{optical and spectral properties of the sample}
All the 557 radio sources identified in the 2dF catalogue are presented in 
Tables 1 and 2 at the end of the Paper. For each object the Tables indicate:

\noindent
(1) Source number\\
(2) Right Ascension $\alpha$ (J2000)
and Declination $\delta$ (J2000). Note that 
these correspond to the 
FIRST radio coordinates, except in the case of objects with double or triple 
sub-structures, where the coordinates are of the centroid of the source and 
are obtained by following the procedure illustrated by Magliocchetti et 
al. (1998) to combine multi-component objects.\\
(3) Offset (expressed in arcsecs) of the optical counterpart in the APM 
catalogue.\\
(4) Radio-flux density (in mJy units) at 1.4 GHz.\\
(5) Apparent $b_J$ and, when present, (6) R magnitudes of the optical 
counterpart.\\
(7) Redshift.\\
(8) Spectral Classification. It is worth remarking here that, as Madgwick et al. 
(2001) have recently shown, the fibre spectral classification inconsistency claimed 
by Kochanek, Pahre \& Falco (2001) is not a problem for 2dFGRS. We refer 
the reader to Madgwick et al. (2001) for further discussion on this point.\\
(9) Emission lines detected ordered from the most to the least prominent. Note 
that the [N${\rm II}$] line appears in the Tables only if its intensity 
was comparable with the H${\alpha}$ one.\\  
(10) Notes on morphological appearance.\\

The point (10) has been obtained by visual inspection of the radio (R, 
from the FIRST atlas) and optical (O, from the Digitized Sky Survey)
images of each source. Blank spaces correspond to point-like
structures as observed both in the radio and optical bands. Radio
images have been classified as {\it Extended\/} whenever it was possible
to detect any extended structure, {\it Double\/} if the source presented
the two characteristic lobes, {\it Core+lobes\/} if also the central
component of the radio source was present and {\it Jets/Core+Jets\/} if
the structures pointing towards the lobes were visible.  In a few
cases, the `?' indicates uncertainty to distinguish between genuine
double/triple structures and the presence of side-lobes.

From the optical point of view, galaxies have been classified 
as {\it Spiral\/} whenever the presence of spiral arms was visible or 
{\it Disk+bulge\/} if the image  clearly resolved the galaxy into 
the two components, {\it Irregular\/} if they presented distorted  
morphologies, {\it Merging\/} if there were signatures of either merging or 
accretion of a smaller unit, and {\it Interaction\/} whenever two or more 
galaxies were interacting with each other. Again, the `?' indicates dubious 
classifications.

\begin{figure*}
\vspace{13cm}  
\includegraphics{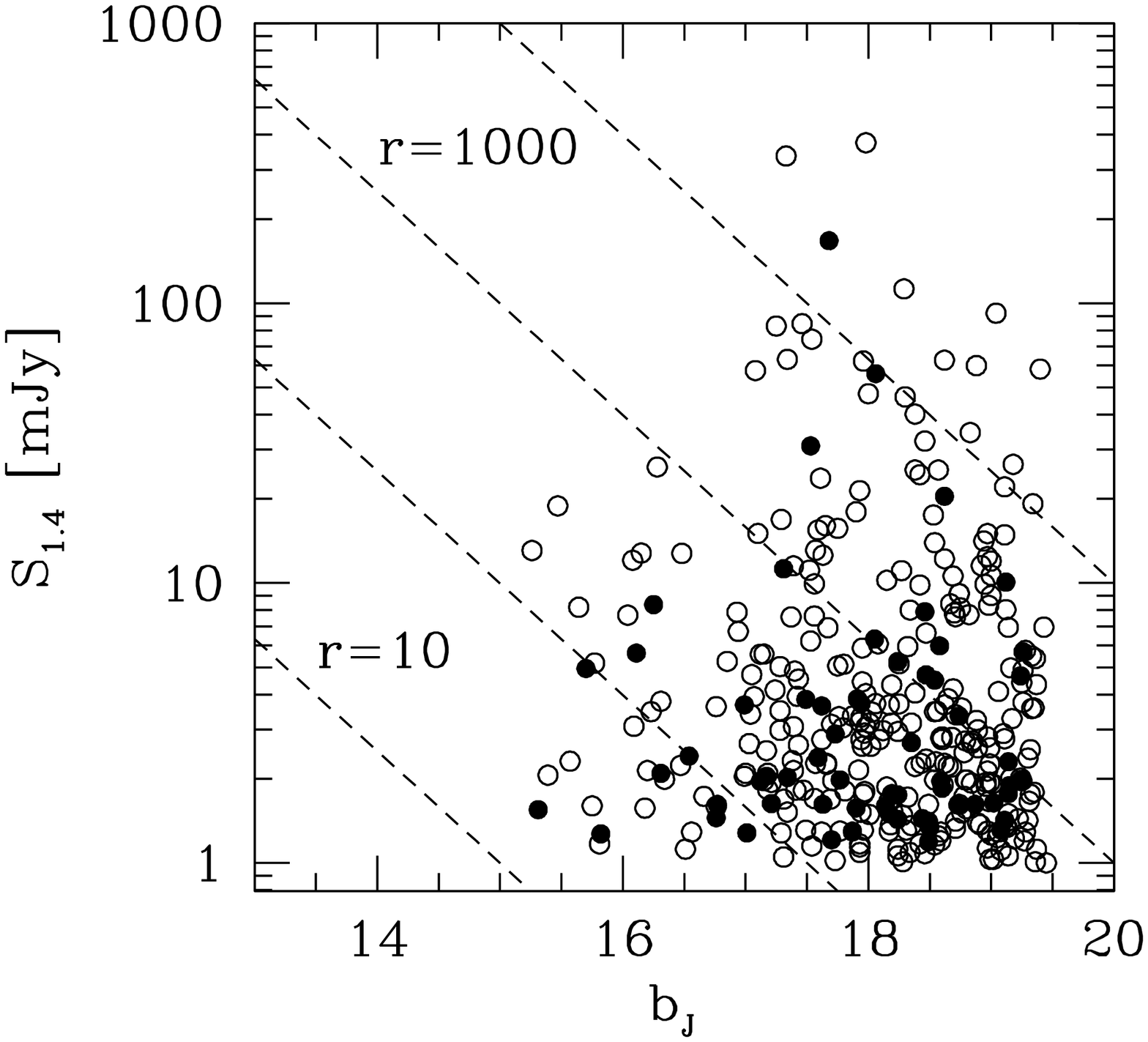}
\includegraphics{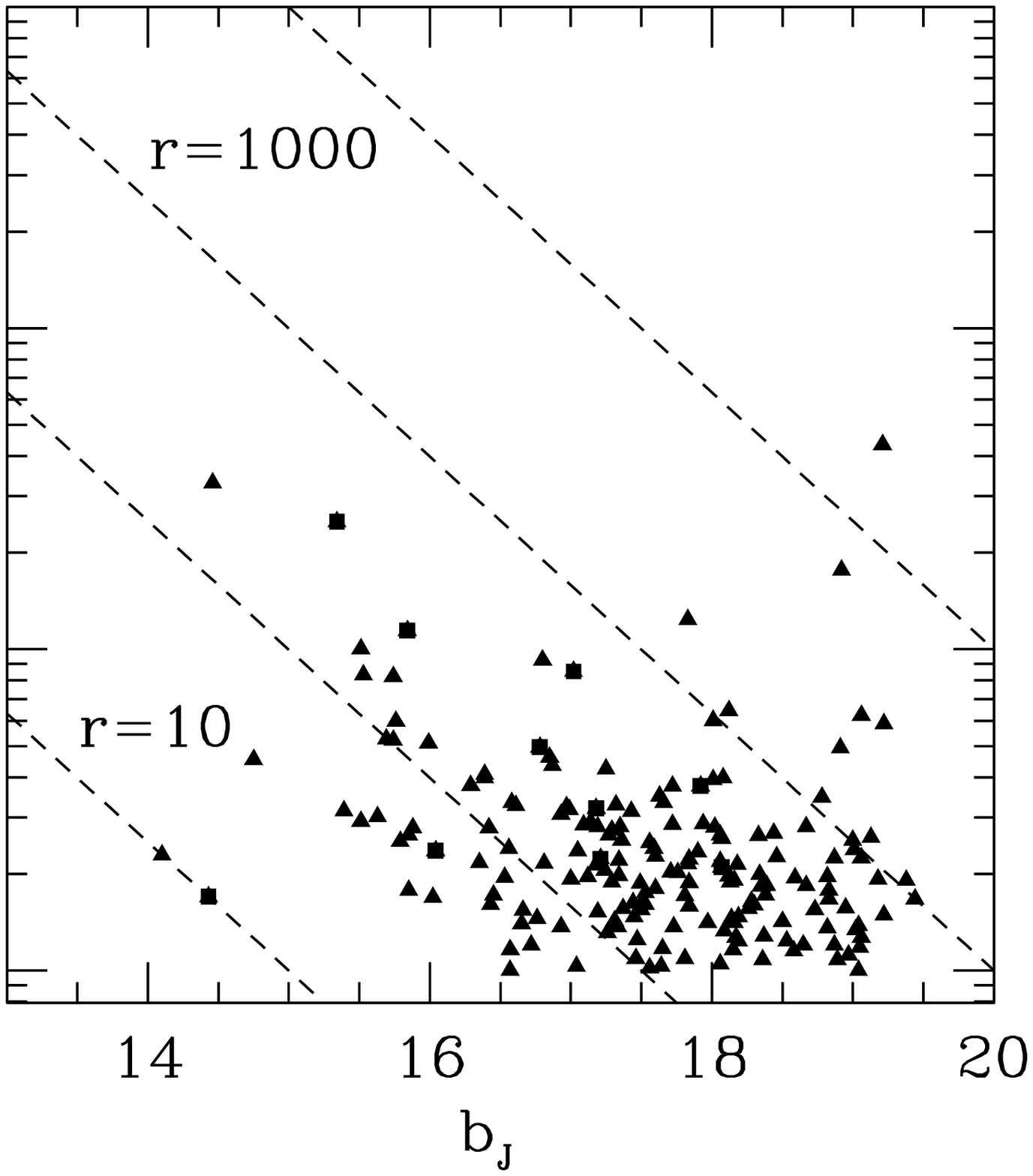}
\includegraphics{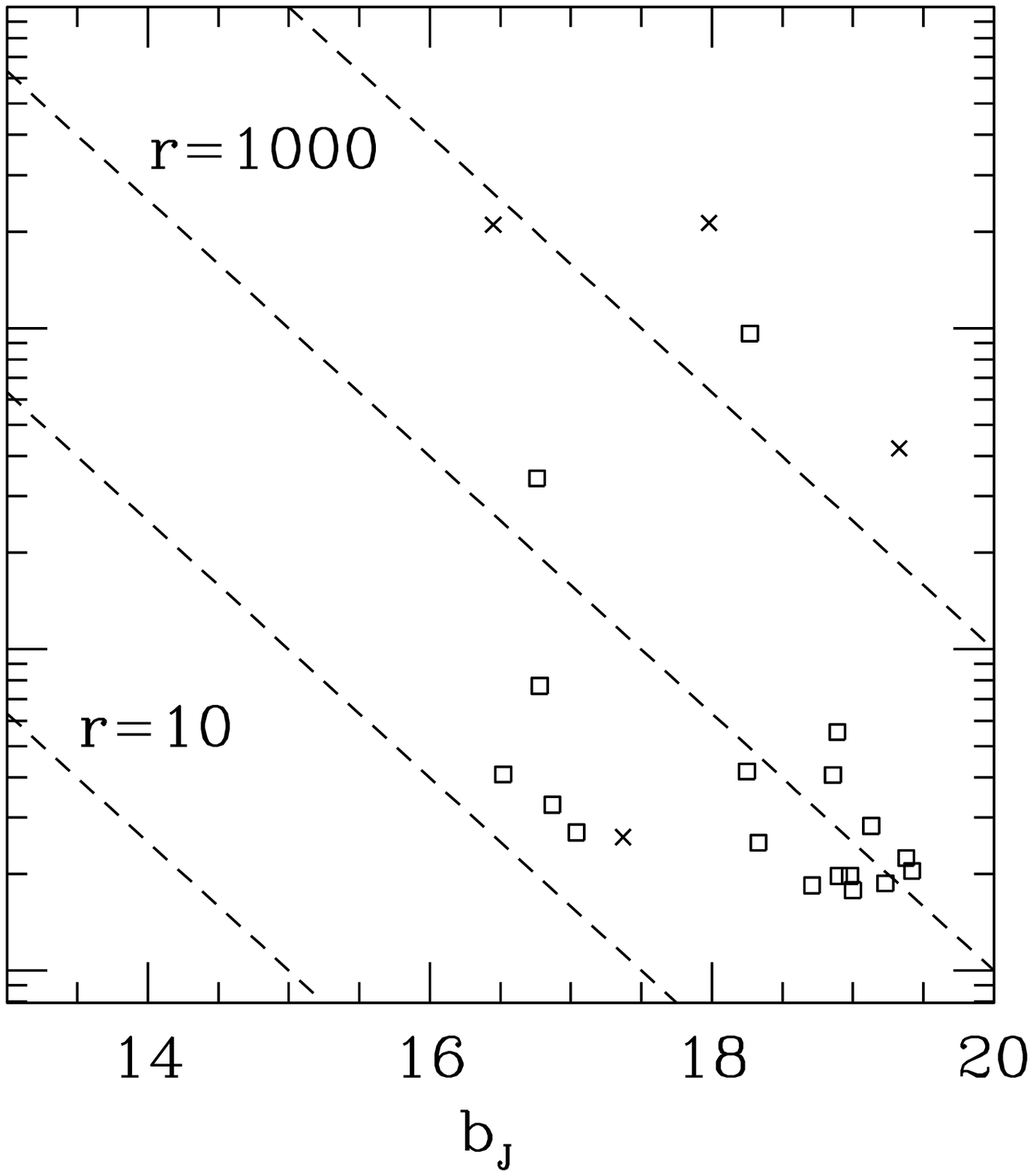}
\includegraphics{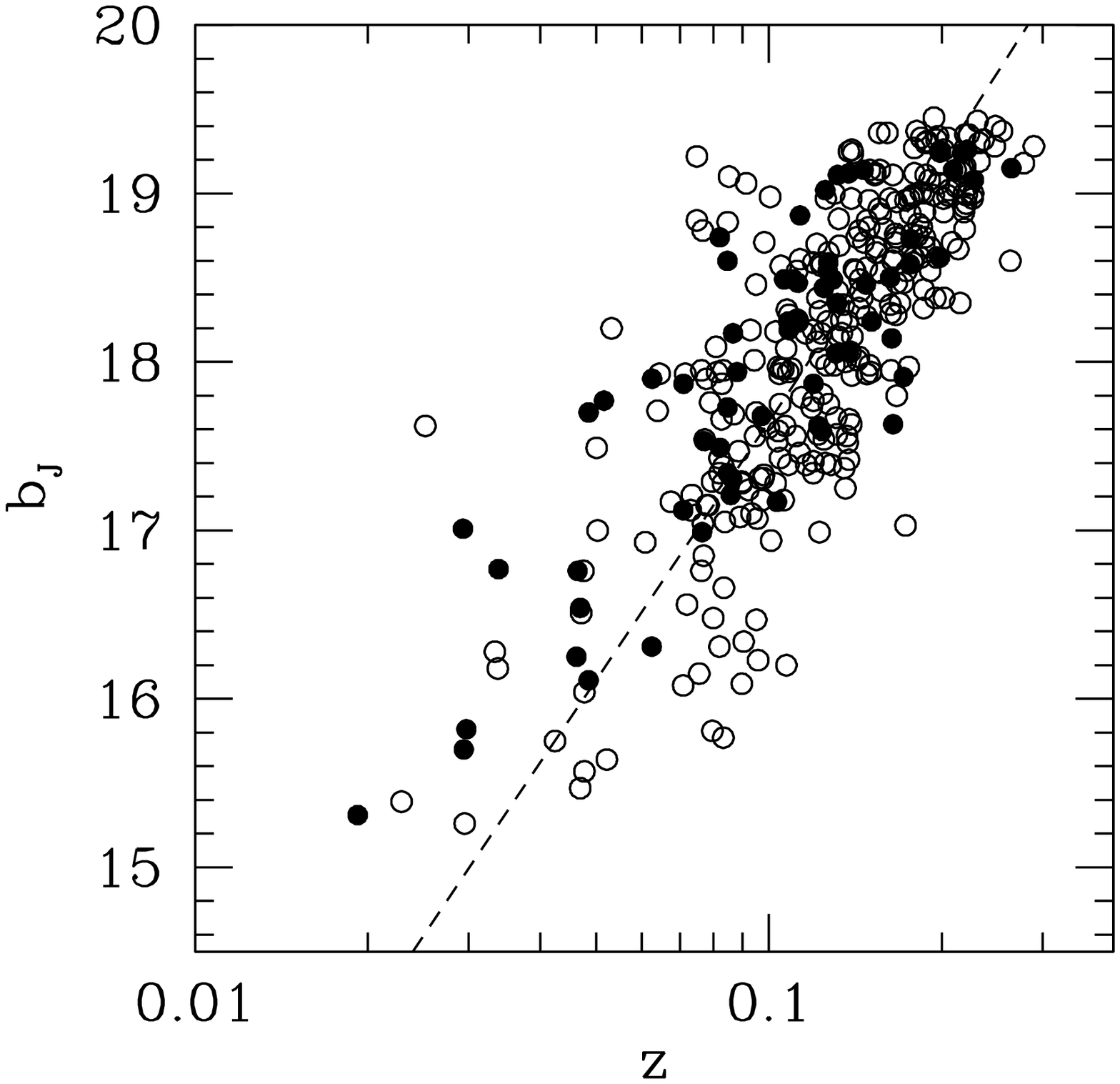}
\includegraphics{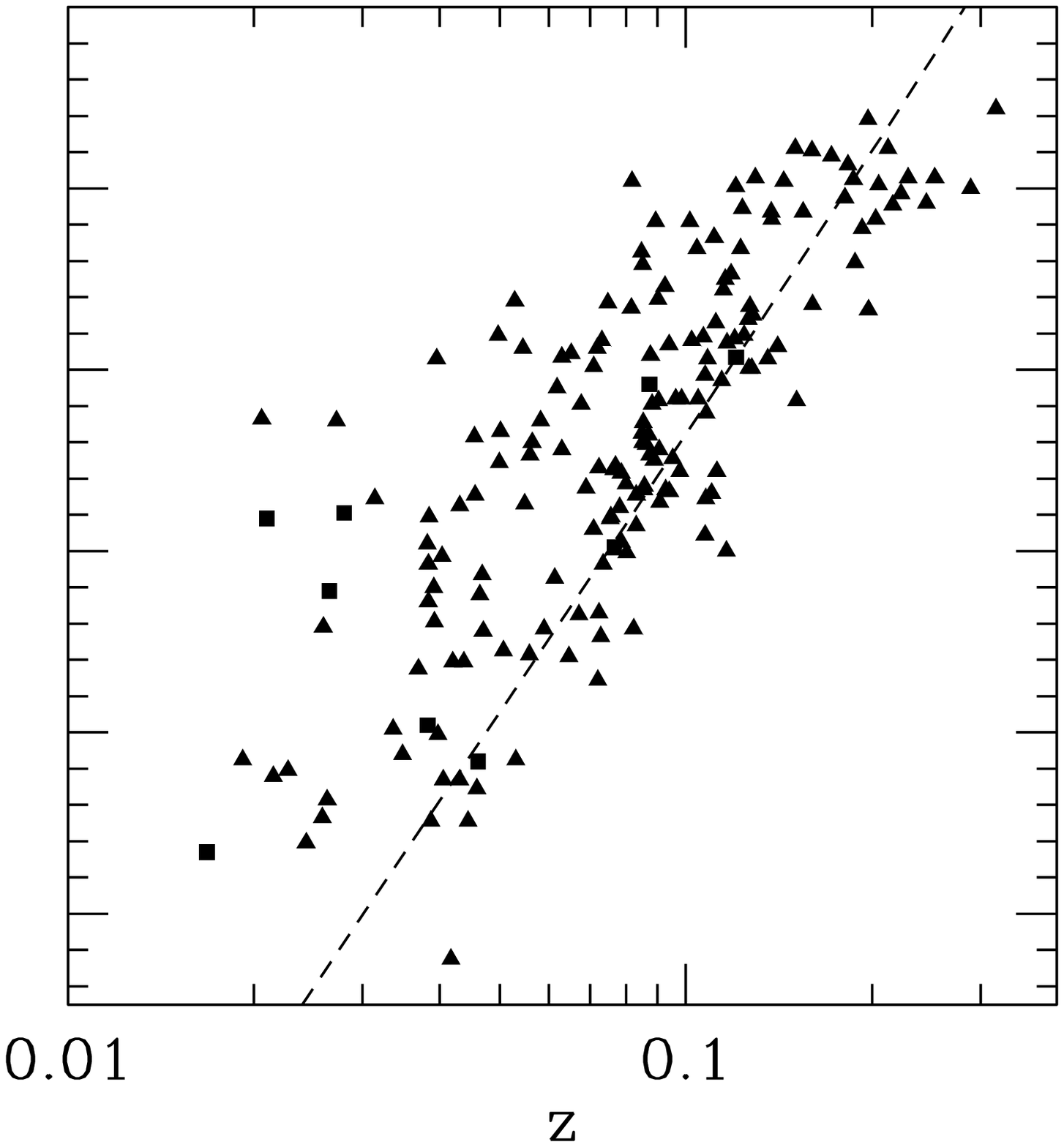}
\includegraphics{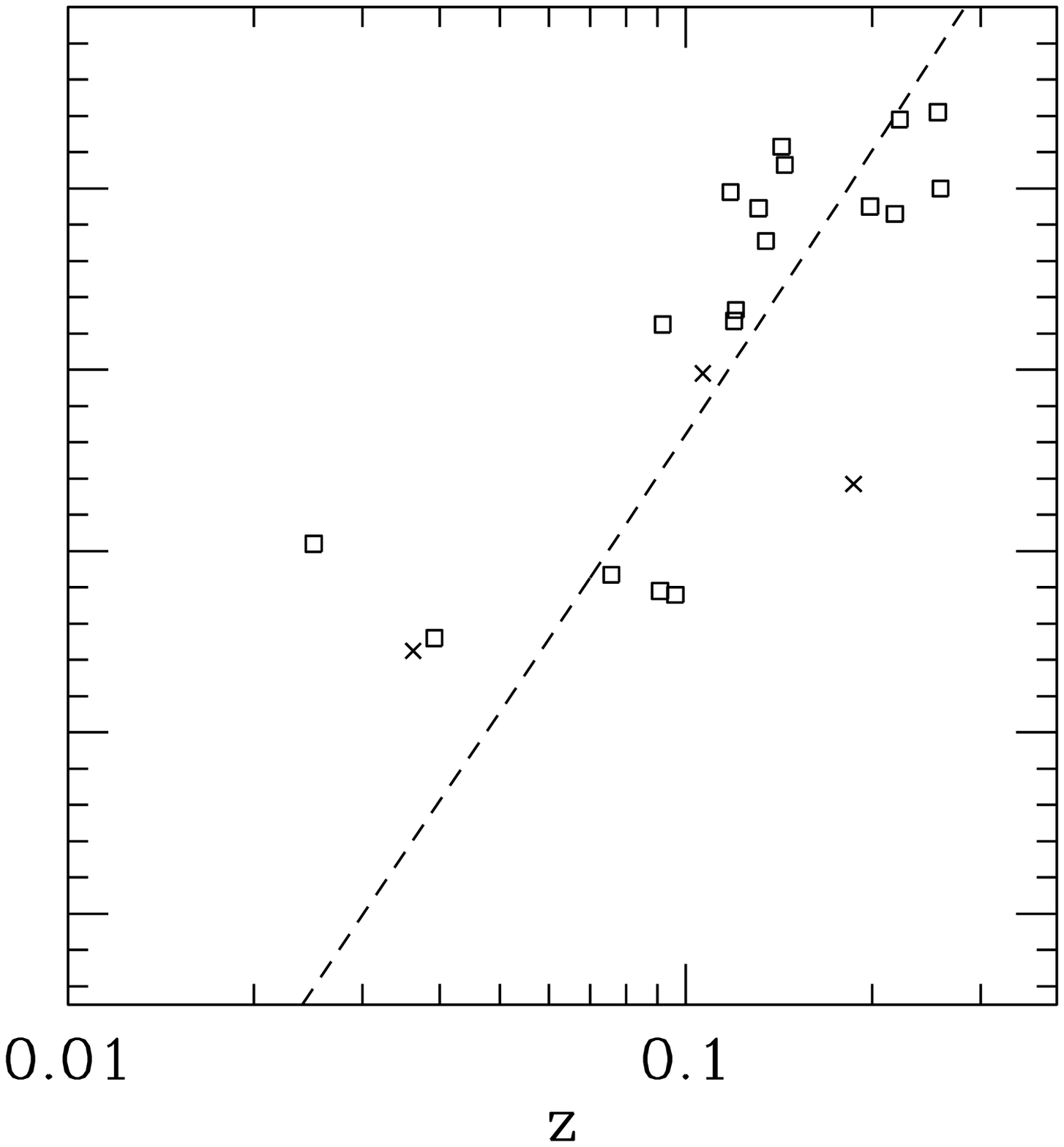}
\caption{$b_J$ magnitudes versus radio flux $S$ at 1.4 GHz (top panels) and 
redshift (bottom panels) for the different classes of objects 
discussed in the Paper. From left to right, plots are derived for early-type 
(represented by empty circles) and E+AGN (filled circles) galaxies, 
late-type (filled triangles) and starburst (filled squares) galaxies and 
Seyfert 1/Seyfert 2 (respectively 
illustrated by crosses and empty squares) objects. The few 
uncertain classification sources are not included in the plots.    
The dashed lines in the top panels correspond to constant values of the 
radio-to-optical ratios $r=1,10,100,10^3, 10^4$, while those in the $b_J-$z 
plots represent the best fit to the data obtained for 
the population of early-type galaxies, corresponding to an absolute magnitude 
$M_B=-21.3$ (see text for details).
\label{fig:F_B}}   
\end{figure*} 

Classes for the optical counterparts of radio sources (column 10 of  
Tables 1 and 2) have been assigned on the basis of their 2dF spectra.
Spectra have been compared with known templates (see e.g. Kennicut 1992; 
McQuade et al. 1995)  which allowed galaxies to be divided into 6 broad 
categories:
\begin{enumerate}
\item {\it Early-type galaxies}, where spectra were dominated by continua much 
stronger than the intensity of any emission line. These objects can be 
furtherly divided into two sub-classes:\\
\hglue\parindent (i)  galaxies with absorption lines only.\\
\hglue\parindent (ii) galaxies with absorption lines + weak [O${\rm II}$] and H$\alpha$ 
     emission lines denoting little star-formation activity.
\item {\it E+AGN-type galaxies}, showing spectra typical of early-types 
plus the presence of (narrow) emission lines such as [O${\rm II}$], 
[O${\rm III}$], [N${\rm II}$] and [S${\rm II}$], which are strong if 
compared to any Balmer line in emission and indicate the presence of large, 
partially ionized transition regions as is the case in active galaxies.  
\item {\it Late-type galaxies}, where spectra show strong emission (mainly 
Balmer) lines characteristic of star-formation activity, together with a 
detectable continuum.
\item {\it Starburst (SB) galaxies}, with optical spectra
characterized  by an almost negligible continuum with very strong 
emission lines indicating the presence of intense star-formation activity. 
\item {\it Seyfert 1 galaxies}, with spectra showing strong, broad emission 
lines.
\item {\it Seyfert 2 galaxies}, where the continuum is missing and spectra 
only show strong narrow emission lines due to the presence of an active 
galactic nucleus.
\end{enumerate}

\begin{figure*}
\vspace{13cm}  
\includegraphics{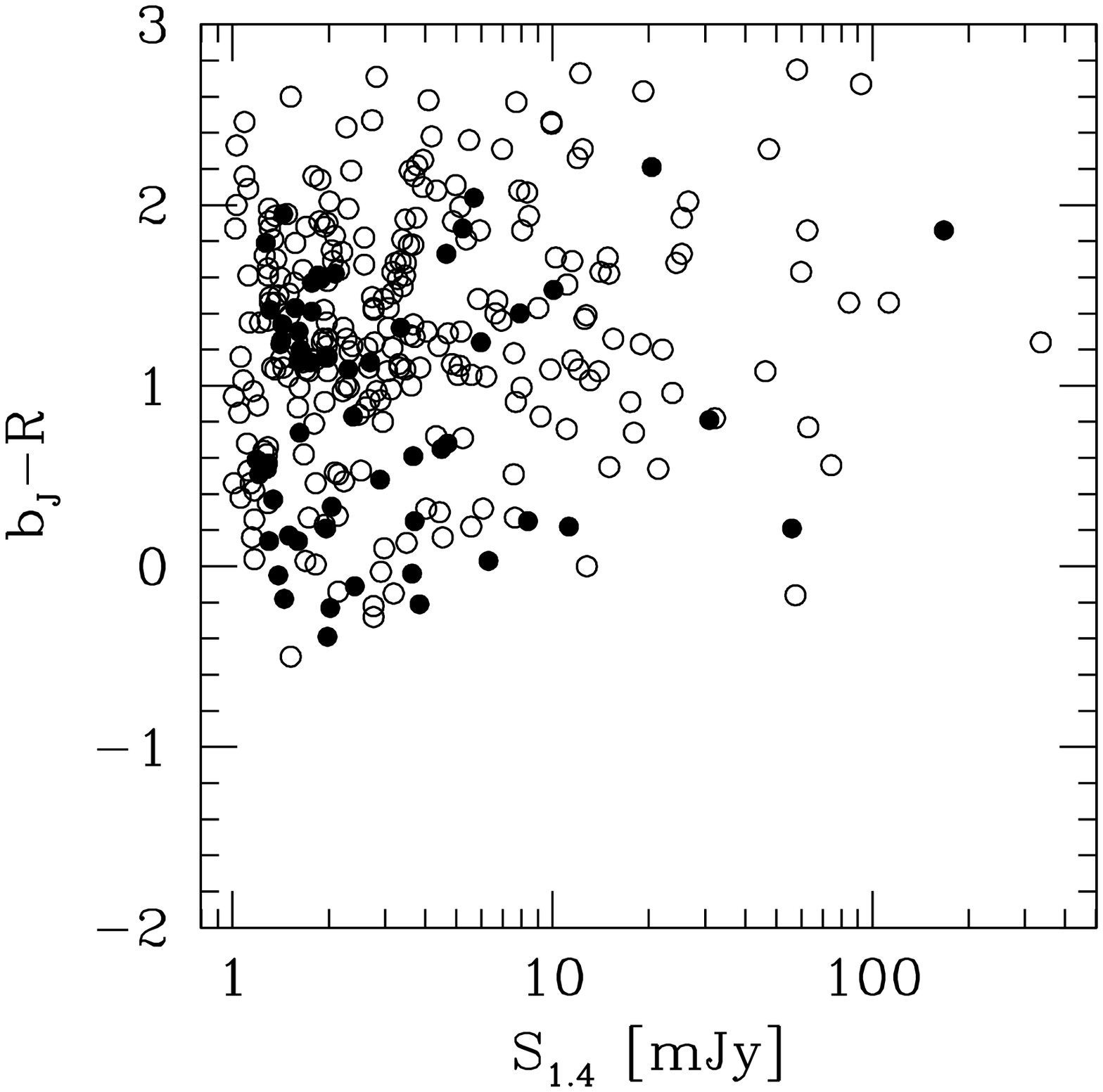}
\includegraphics{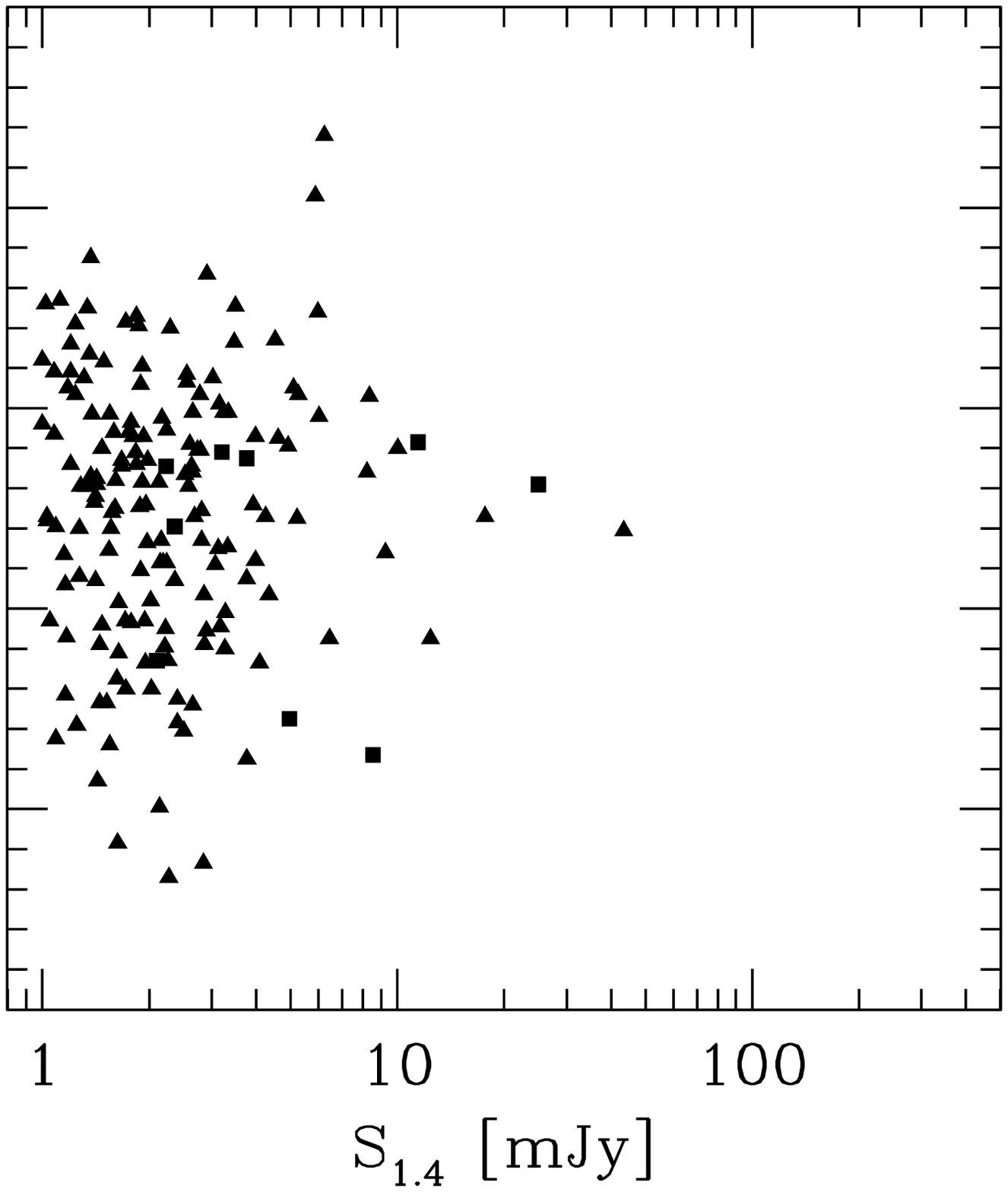}
\includegraphics{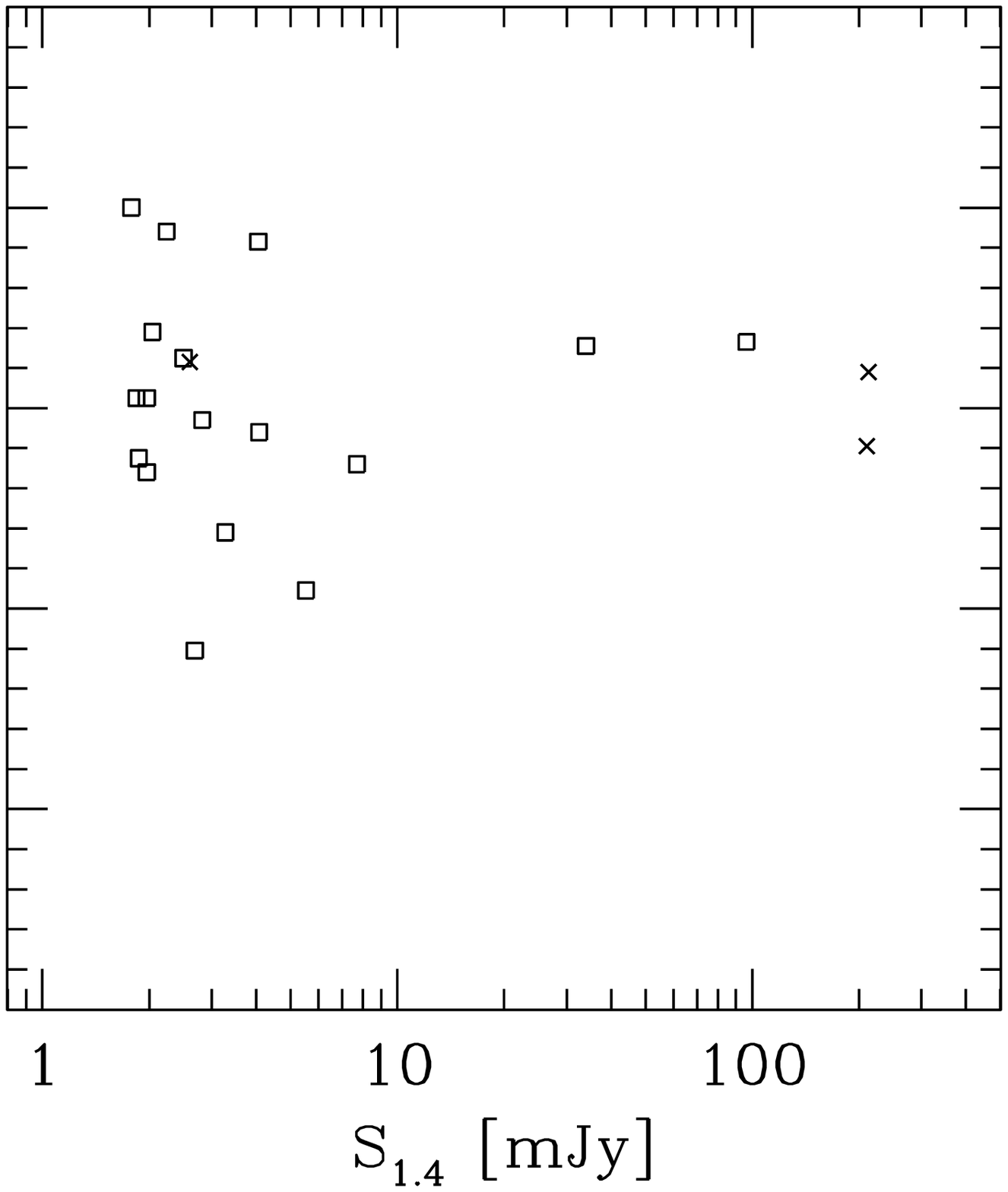}
\includegraphics{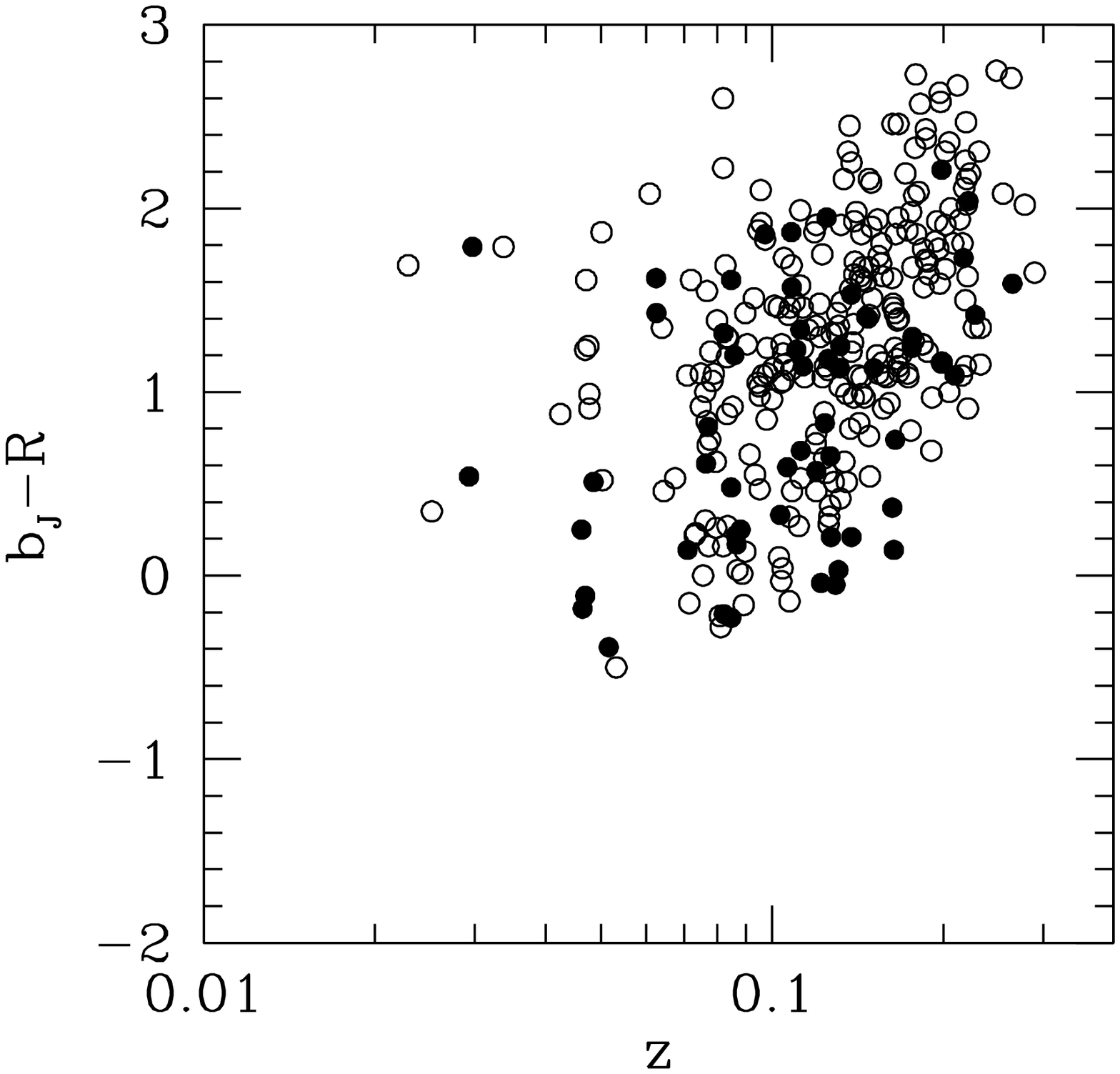}
\includegraphics{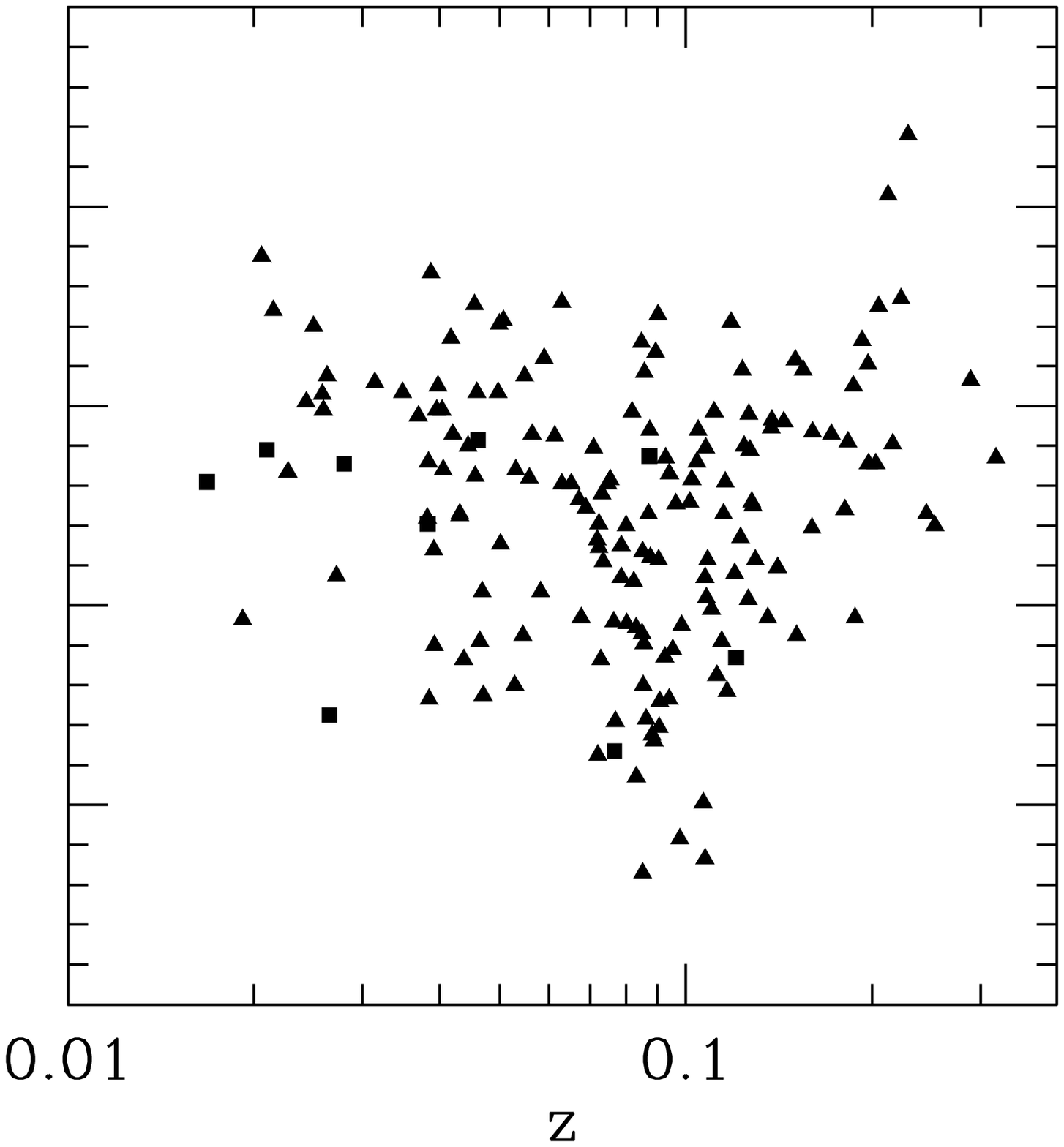}
\includegraphics{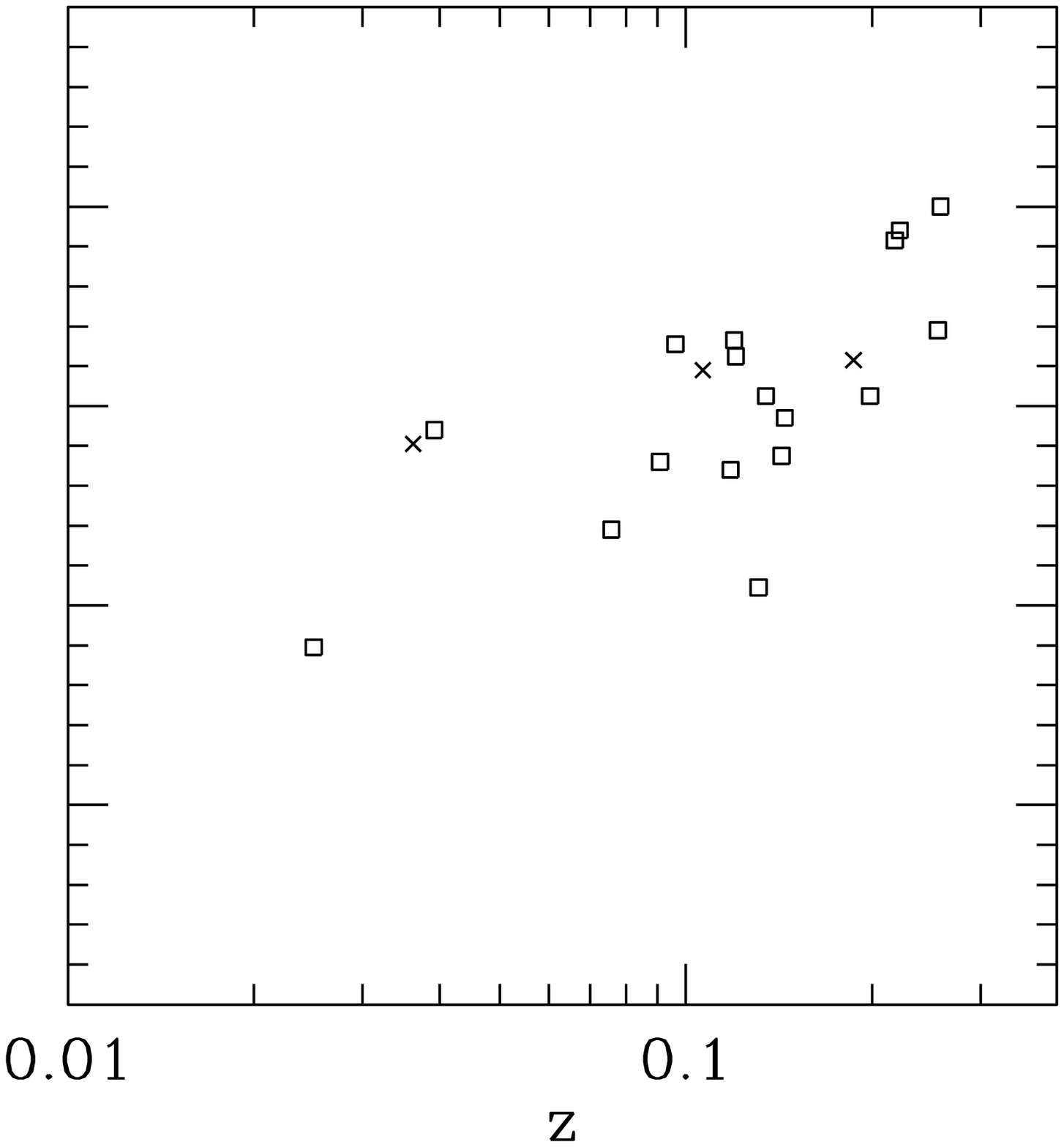}
\caption{Upper panel: $b_J-R$ colours for the different classes of sources 
as a function of radio flux at 1.4 GHz, $S$. Bottom panel: $b_J$-R colours 
as a function of redshift. Empty circles represent early-type galaxies, 
filled circles E+AGN, filled triangles are for late-type galaxies, filled 
squares for starbursts, empty squares for Seyfert 2 and crosses illustrate 
the case of Seyfert 1 objects. 
\label{fig:BR_S}}   
\end{figure*} 

Lastly, we also found 3 galactic stars, probably due to 
random positional coincidences with extragalactic radio sources.

Distinctions amongst different classes of sources and in particular
amongst E+AGN, Seyfert 2 and Late-type galaxies have relied on the
diagnostic emission line ratios of Veilleux \& Osterbrock (1987),
Waltjer (1990) and Rola Terlevich \& Terlevich (1997) (where this last 
approach, based on the relative intensity of H$\beta$, has been adopted 
whenever the H$\alpha$ line would fall off the spectrum, i.e. for 
$z\simgt0.2$). Note that a
definite classification was not possible for all the cases. This
simply reflects the fact that it is in general quite common to find
`composite' galaxies containing both an AGN and ongoing star
formation (see e.g. Hill et al. 2001).

The photometric properties of the different classes which constitute
the spectroscopic sample are illustrated by Figs 5 and 6. In all
the plots, symbols are as follows: empty circles for early-type
galaxies, filled ones for E+AGN, filled triangles for late-type
galaxies, filled squares for starbursts, empty squares for Seyfert 2
and crosses for Seyfert 1 objects. Note that we did not include here
the small number of uncertain classification sources.

Fig. 5 respectively shows $b_J$ magnitudes versus radio flux $S$ at 1.4~GHz 
(top panels) and redshift (bottom panels) for the different classes of 
objects in the spectroscopic sample; left-hand panels are for early and 
E+AGN galaxies, middle ones for 
late-type and starburst galaxies, while the right-hand panels represent the 
populations of Seyfert 1 and Seyfert 2 galaxies. 
The dashed lines (in the top panels) indicate the loci of constant 
radio-to-optical ratios $r=S\times 10^{(m-12.5)/2.5}$, where S is expressed 
in mJy and $m\equiv b_J$. Note that $R-$z and $R-S$ trends for the different 
classes of sources are not presented here, as they simply reproduce the 
same features already shown in Fig. 5.
Figs 6 instead represents $b_J-R$ colours for the different populations 
as a function of radio flux $S$ (top panels) and redshift (bottom panels).
Symbols are as in Fig. 5.

As one can notice from Figs 5 and 6, different classes of objects 
occupy different regions on the various  $S-b_J$, $b_J$-z, etc. planes; 
this is due to intrinsic characteristics of the various populations 
which we summarize here:\\
\strut\\
(1) Early-type galaxies.\\
This class comprises 289 objects and makes up 52 per cent of the whole
spectroscopic sample. On average, 
sources belonging to this population have relatively high
values for the radio to optical ratio ($100\simlt r\simlt 10^4$) and are
preferentially found at redshifts $z \simgt 0.1$. 
The majority of these sources show rather red colours
($b_J-R\simgt 1$), with a tendency to be redder at larger look-back
times. Their radio fluxes lie in the range $1\simlt S/{\rm mJy}
\simlt 10$, even though objects are found up to $S\sim 400$~mJy, and
optically they appear as relatively faint (about 60 per cent of the
sources has $b_J\ge 18$).  The $b_J-$z relation in this case
(represented by the dashed line in the bottom-left panel of Fig. 5)
is well described by the relation
\begin{eqnarray}
b_J-M_B=-5+ 5\;{\rm log_{10}}d_L({\rm pc})\simeq
25-5{\rm log}_{10}{\rm H}_0+\nonumber \\+5{\rm log}cz+1.086 (1-q_0)z
\end{eqnarray}
($d_L$ is the luminosity distance, $q_0=\Omega_0/2$ and $M_B$ is the absolute magnitude
in the $b_J$-band). A $\chi^2$ fit to the data performed with $M_B$ as
free parameter gives, for an $h_0=0.5$, $\Omega_0=1$ universe,
$M_B= -21.29$ with little scatter about this value ($\Delta
M_B= 0.28$~mag).  This result is in very good agreement with
previous estimates (see e.g. Rixon, Benn \& Wall 1991; Gruppioni et
al. 1999; Georgakakis et al. 1999; Magliocchetti et al. 2000),
showing that passive radio galaxies are reliable standard candles.\\
\strut\\
(2) E+AGN galaxies.\\
There are 61 objects in this class, corresponding to 11 per cent of
the spectroscopic sample.  These sources are directly connected to the
class of early-type galaxies, even though they show characteristics
that are intermediate between pure AGN-fuelled sources and
star-forming galaxies. Optically they appear as rather faint and
closely follow the standard-candle relationship found for early-type
galaxies ($M_B=-21.45\pm 0.41$~mag).  
However their radio-to-optical ratios are in general as low
as those obtained for late-type galaxies.  Their radio fluxes are in
general quite low ($S\simlt 5$~mJy) and their $b_J-R$ colours are
uniformly distributed between about 0 and 2.\\
\strut\\
(3) Late-type galaxies and starbursts.\\
This class comprises 177 objects (including 10 SB), which is $\sim 30$
per cent of the spectroscopic sample.  The radio-to-optical ratios $r$
for these sources have values between 60 and 10$^3$, in general
smaller than that found for early-type galaxies. Their radio fluxes
are rarely brighter than $S\simeq 5$~mJy and optically they have
bright apparent magnitudes (very few objects are found with $b_J\simgt
18.5$).  These sources have quite blue colours, $-2\simlt b_J-R \simlt
1$, and are mostly local -- the majority of them being found within
$z \simeq 0.1$. Furthermore, 
do not follow the $b_J-$z relation
found for early-type galaxies, illustrated by the dashed line in the
middle panel at the bottom of Fig. 5. For these objects one in fact finds 
$M_B=-20.76\pm 0.33$~mag.\\
\strut\\
(4) Seyfert galaxies.\\
The sample contains 18 Seyfert 2 and 4 Seyfert 1 galaxies, making up
$\sim 3$ and $0.5$ per cent of the sample respectively. Their
radio-to-optical ratios are quite high, and in the case of all but one
Seyfert 1 galaxy, one finds $r\simgt 10^4$. Their colours range
between $0\simlt b_J-R \simlt 2$ as in the case for E+AGN galaxies,
and they seem to follow the $b_J-$z relationship found for early-type
galaxies (bottom-left panel of Fig. 5).  Seyfert 2 galaxies in
general have low-to-intermediate radio fluxes, while Seyfert 1's show
higher values for $S$. Note that one Seyfert 1 galaxy (number 331\,437
of the spectroscopic sample) is found at $z=2.24$, well beyond the
typical limit of the 2dFGRS, $z\simeq 0.3$.  

The above findings for the relative contribution of the different
classes of sources to the local radio population are in good
general agreement with the results of Sadler et al. (2002) for a sample
of NVSS/2dFGRS galaxies. However, Sadler et al. (2002) find approximately
40 per cent are starforming galaxies, the increased fraction almost
certainly being due to the lower-resolution NVSS survey which will detect
(or catalogue) low surface brightness sources at the lowest flux densities.


Some information on the above classes of sources can also be derived from the 
few optical images showing resolved structures. For instance, it is 
interesting to note that the majority of the interacting/merging systems seem to be
associated with early-type spectra, typical of pure AGN-fuelled sources, 
suggesting that merging, under appropriate conditions, can trigger AGN 
activity even at low redshifts. Also, as expected, irregulars and spirals 
preferentially show spectra typical of late-type galaxies; signatures 
of interaction and/or merging are seen for members of this latter population 
as well as for E+AGN galaxies. Interaction/merging is also observed for one 
of the four Seyfert 1 galaxies and two Seyfert 2's. Furthermore we find one 
Seyfert 2 to be associated to the classical two-lobed radio image.

Note that in a few cases, early-type/E+AGN galaxies are found to
correspond to spiral morphologies. This is simply due to our
classification criteria which is based solely on the spectra; 
typical spectra of Sab spiral galaxies can be very similar to
early-type galaxies with [O${\rm II}$] and H${\alpha}$ emission
lines.  Finally, radio images show that there is a clear trend for
extended/sub-structured sources to be associated with absorption
systems (i.e. early-type galaxies).

\section{radio properties of the sample}


\begin{figure}
\vspace{8cm}  
\includegraphics{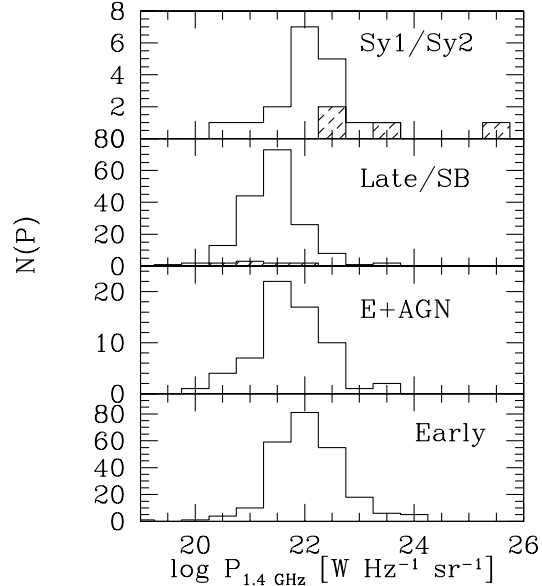}
\caption{Distribution of monochromatic radio power at 1.4 GHz for the
6 classes of sources discussed in the paper. In the first and second panels 
from top, shaded histograms respectively illustrate the distribution of 
Seyfert 1 galaxies and starbursts (not clearly seen due to the small number 
of sources).    
\label{fig:power_hist_type}}   
\end{figure}

More information on the nature of the sources in the spectroscopic sample can 
be inferred by investigating radio properties such as the 
distribution of their radio powers and the radio luminosity function 
(hereafter denoted as LF). 

From investigation of the radio-flux distribution as a function of
redshift for the different populations as inferred from Fig.
5 (top and bottom panels), we note that in general Seyfert 1 and early-type
galaxies show higher radio fluxes compared to the other classes of
sources.  There is also a tendency for late-type and SB galaxies to be
only present in the local universe, as already mentioned.  No other
distinction amongst the different populations is evident in the data.

Radio powers have been calculated according to the relation $P_{1.4\rm GHz}=
S_{1.4\rm GHz} D^2 (1+z)^{3+\alpha}$, and are expressed in $\whzsr$ 
units. In the above formula, $D$ is the angular
diameter distance and $\alpha$ is the spectral index of the radio
emission ($S(\nu)\propto \nu^{-\alpha}$). As we did not have measured
values for this latter quantity, we assumed $\alpha=0.5$ for Seyfert 1
galaxies, $\alpha=0.75$ for early-type galaxies (with or
without emission lines due to AGN activity), $\alpha=0.7$ for Seyfert 2's and 
$\alpha=0.35$ both for late-type galaxies and starbursts (see e.g. Oort et al. 
1987). Note however that, since the redshift range is relatively small, 
the results do not depend on the precise values of the spectral index. 
Fig. \ref{fig:power_hist_type} shows the distribution of radio 
luminosities 
for the different classes of sources. In the top panel, the shaded histogram 
corresponds to the distribution of Seyfert 1 galaxies, while in the second 
panel from top it represents the population of SB galaxies (not very clearly 
seen due to the small number of sources). As one can see from the Figure, 
early-type galaxies show in general higher radio powers 
($10^{21}\simlt P_{1.4\rm GHz} / \whzsr \simlt 10^{23}$, 
extending up to $10^{24} \whzsr$)
than late-type ones, whose majority presents $P_{1.4\rm GHz}\simlt 10^{21.5} \whzsr$. 
E+AGN sources are instead characterized by luminosities with values 
closer to those found for early-type galaxies 
($10^{20}\simlt P_{1.4\rm GHz}\whzsr \simlt 10^{22.5}$). 
The bulk of the Seyfert 2 distribution is 
obtained for $P_{1.4\rm GHz}\simeq 10^{22} \whzsr$, while 
Seyfert 1 galaxies all have $P_{1.4\rm GHz}\simgt 10^{22.5}\whzsr$. 
We note that the apparent cut in the luminosity 
distribution at $P_{1.4\rm GHz}\simeq 10^{20} \whzsr$ is simply due to the 1~mJy 
limit adopted for sources in the original radio sample ($S=1$~mJy 
corresponds to $P_{1.4\rm GHz}\simeq 10^{20} \whzsr$ for 
$z=0.02$, redshift of the nearest object in the spectroscopic sample).

It is worth remarking here that all the objects included in the spectroscopic 
sample (apart from the one Seyfert 1 found at $z=2.24$), have $P_{1.4\rm GHz} 
\simlt 10^{24} \whzsr$. In the case of ``classical'' radio 
sources (i.e. if one excludes late-type galaxies and starbursts from
the analysis, since the radio signal stems from non-AGN activity) this
implies the overwhelming majority of early and E+AGN galaxies to
belong to the class of FR~I sources (Fanaroff \& Riley 1974). In fact,
although the distinction between FR~I and FR~II sources seems to be
more complicated than just based on their radio power -- as in general
optically brighter FR~I also tend to appear as more luminous in the
radio band (see e.g.  Ledlow \& Owen 1996) -- we note that for
$P_{1.4\rm GHz}\simlt 10^{24} \whzsr$ only members of
FR~I are found, independent of their magnitude.

Also, from a morphological point of view, we find that very few
objects in the sample (about 5 percent) show composite structures such
as lobes or jets.  The angular resolution of the FIRST survey is about 
5 arcsec which, for the cosmology adopted in this Paper, 
corresponds to a physical scale of $\sim 7$~kpc at $z=0.05$ and $\sim 35$~kpc 
at $z=0.3$ (the maximum redshift found for galaxies in the 2dFGRS).    
Thus for all the sources showing point-like images in the
radio, one can exclude the presence of very extended structures such
as those typical of FR II sources.

The local radio LF for objects in the spectroscopic sample has then been 
derived by grouping the sources in bins 
of width $\Delta {\rm log_{10}} P=0.4$, according to the expression
\begin{eqnarray}
\Phi(P)=\sum_i N_i(P, P+\Delta P)/V_{\rm max}^i(P),
\label{eq:phi}
\end{eqnarray}
where $N_i$ is the number of objects with luminosities between $P$ and
$P+\Delta P$ and $V_{\rm max}^i(P)$ is the maximum comoving volume within
which an object could have been detected above the radio-flux and
magnitude limit of the survey (Rowan-Robinson 1968). 
As already stated in Section 2, for the radio-flux limit we set $S=1$~mJy, 
value at which the incompleteness of the original radio catalogue has been 
well assessed to be $\sim 20 \%$, while $b_J=19.45$ -- derived for 2dFGRS 
galaxies -- is the chosen magnitude limit. Note that this value could be 
transformed into a rough redshift estimate for completeness of the 
spectroscopic sample, $z\simeq 0.2$, under the assumption of radio galaxies as 
standard candles with $M_B\simeq -21.3$ (see equation 1 and Fig. 5).

For each source we then estimated the maximum redshift at which it could have 
been included in the sample, $z_{\rm max}={\rm min}(z_{\rm max}^R, z_{\rm max}^O)$, where 
$z_{\rm max}^R$ and $z_{\rm max}^O$ are the redshifts at which the source 
would disappear from the sample respectively due to radio and optical limiting 
flux densities. For the optical K-corrections we adopted, under the 
assumption of spectra of the form $S\propto \nu^{-\alpha_{\rm opt}}$, 
$\alpha_{\rm opt}=5$ for early-type galaxies, $\alpha_{\rm opt}=3$ for 
late-type galaxies and $\alpha_{\rm opt}=0.5$ for Seyferts (both type 1 
and 2; see Rowan-Robinson 1993). 

In principle, in order to derive the local LF, one should de-evolve the 
luminosity of each source at $ z=0$ according to expressions such as 
$P(0)=P(z)\;\rm exp^{-t(z)/\tau}$ (where in this case we assumed pure 
luminosity evolution), where $t(z)$ is is the look-back time in units of the 
Hubble time and $\tau$ is the time-scale of the evolution in the same units.
However, given the limited redshift range spanned by the sources in the 
spectroscopic sample, we assumed no evolution (i.e. $\tau\rightarrow \infty$) 
for all the classes of objects and therefore set $P(0)\equiv P(z)$. 

\begin{figure}
\vspace{8cm}  
\includegraphics{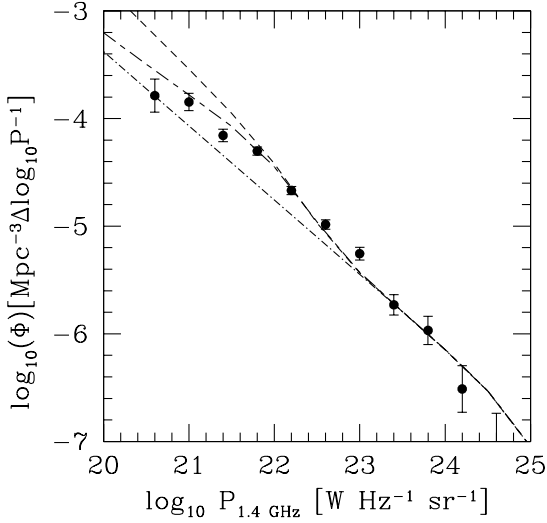}
\caption{Local radio luminosity function at 1.4 GHz for all the objects in 
the spectroscopic sample with $S\ge 1$~mJy and $b_J\le 19.45$. The dashed 
line indicates the prediction from Rowan-Robinson et al. (1993), while the 
dashed-dotted line 
represents the predicted LF for the population of AGN-fuelled steep spectrum 
sources as given by Dunlop \& Peacock (1990) under the hypothesis of pure 
luminosity evolution. The long/short-dashed line is obtained from equations 
(3), (\ref{eq:phiell}) and (\ref{eq:phisp})
for $P_*=10^{21.7} \whzsr$, $\beta=1.4$, $C=0.425\times 
10^{-4}$ and $\sigma=0.44$ (see text for details).   
\label{fig:L_z}}   
\end{figure}

\begin{figure}
\vspace{8cm}  
\includegraphics{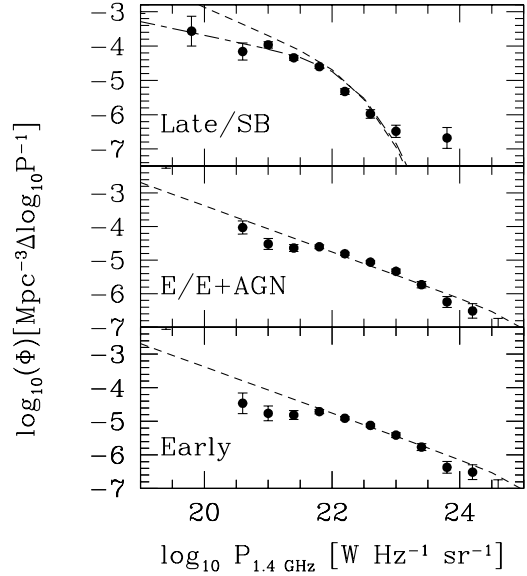}
\caption{Local radio luminosity function at 1.4 GHz for late-type+SB galaxies 
(top panel), early and E+AGN galaxies (middle panel) and early-type 
galaxies only (bottom panel). All the sources have $b_J\le 19.45$, 
completeness limit of the spectroscopic sample, and $S\ge 1$~mJy. The dashed 
lines in the middle and bottom panels indicate the Dunlop \&
Peacock (1990) model for pure luminosity evolution for the population of 
AGN-fuelled, steep-spectrum sources, while the one in the top panel represents 
the prediction from Rowan-Robinson et al. (1993) for spirals/starburst 
galaxies. The long/short-dashed line is obtained from equation (\ref{eq:phisp})
for $P_*=10^{21.7} \whzsr$, $\beta=1.4$, $C=0.425\times 
10^{-4}$ and $\sigma=0.44$ (see text for details). Seyfert 1/2 galaxies are not 
represented due to their small number.
\label{fig:lum_fcntype}}   
\end{figure}

The values for the LF obtained as in equation (\ref{eq:phi}) have then been 
corrected respectively by means of the factor $0.8\times 556/971$ to take into account 
the 80 per cent completeness of the FIRST survey for sources between 1 and 2~mJy, 
$0.95\times 556/971$ for sources between 2 and 3~mJy and simply 556/971 for sources
brighter than 3~mJy (see Becker et al., 1995), where the ratio 556/971 indicates 
the fraction of objects with $b_J\le 19.45$ observed by the 
2dFGRS in the considered area (see Section 2.3), and reported in Table~1 (first 
column).

Fig. \ref{fig:L_z} shows the results for the whole spectroscopic
sample, with error bars showing the Poisson errors.  The dashed line
indicates the prediction from Rowan-Robinson et al. (1993) who
consider a luminosity function of the form:
\begin{eqnarray}
\Phi(P,z)=\Phi_{\rm SP}(P)+ \Phi_{\rm ELL}(P,z),         
\label{eqn:1}
\end{eqnarray}
where
\begin{eqnarray}
\Phi_{\rm ELL}(P,z)={10^{-6.91} \over
\left[P_{1.4}/P_{\rm c}(z)\right]^{0.69}+
\left[P_{1.4}/P_{\rm c}(z)\right]^{2.17}}
\label{eq:phiell}
\end{eqnarray}
(illustrated in Fig. \ref{fig:L_z} by the dashed-dotted line) is the 
LF for steep spectrum FR~I+FR~II sources as given by Dunlop \& Peacock (1990) 
under the assumption of pure luminosity evolution. The evolving `break'
luminosity is given by ${\rm log_{10}}P_{\rm c}(z)=25.12+1.26 z-0.26 z^2$ (in 
$\whzsr$ units) which, since we have assumed no evolution, reduces to the value corresponding to 
$z=0$, and
\begin{eqnarray}
\Phi_{\rm SP}(P)=C\;\left(\frac{P}{P_*}\right)^{1-\beta}{\rm exp}
\left(-
{
\left[{\rm log_{10}}\left(1+\frac{P}{P_*}\right)\right]^2
\over
2\sigma^2
}
\right),
\label{eq:phisp}
\end{eqnarray}
with $C=0.425\times 10^{-4}$, $\log_{10}P_*=21.83\footnote{This value has
been obtained after correcting the typing error in Rowan-Robinson et
al. (1993; Rowan-Robinson, private communication).} \whzsr$,
$\beta=1.82$, $\sigma=0.44$, is the luminosity function for the
population of spirals/starbursting galaxies as derived for IRAS
sources by Saunders et al. (1990). Note that we also considered a
$\Phi_{\rm SP}$ of the form given by Dunlop \& Peacock (1990), even
though we do not show its trend in Fig. \ref{fig:L_z} given the
unrealistic overprediction of the number of sources for $P\simlt
10^{23} \whzsr$.\\

We note that none of the models can accurately describe the data.
The Dunlop \& Peacock (1990) predictions for radio galaxies underestimate the 
number density of objects for $P\simlt 10^{22.5} \whzsr$, 
while the Rowan-Robinson et al. (1993) model, although able to reproduce the 
departure of the observed LF from the power-law behaviour for 
$P\simlt 10^{23} \whzsr$, overpredicts the number density 
of sources with faint luminosities.

\begin{table*}
\begin{center}
\caption{The local radio luminosity function at 1.4 GHz in Mpc$^{-1}$ 
$\Delta {\rm log_{10}P}$ units. Luminosities are expressed in $\whzsr$.}
\begin{tabular}{lllllllllll}
\noalign{\hrule}
\noalign{\vglue 0.2em}
&$\;\;\;\;\;\;$ Complete sample&&$\;\;\;\;\;\;$ Early-type&&$\;\;\;\;\;\;$E/E+AGN &&$\;\;\;\;\;\;$ Late-type\\           
log$_{10}$P & log$_{10}\Phi$&N$_{\rm All}$& log$_{10}\Phi$&N$_{\rm Early}$& log$_{10}\Phi$&N$_{\rm E/E+AGN}$& log$_{10}\Phi$&N$_{\rm Late}$&\\
\noalign{\vglue 0.2em}
\noalign{\hrule}
\noalign{\vglue 0.2em}
  20.6 &$-3.79\pm0.15$ &8& $-4.46\pm0.30$ & 2& $-4.03\pm0.19$& 5&$-4.16\pm0.25$&3  \\
  21.0 &$-3.85\pm0.08$ &29& $-4.76\pm0.21$ & 4& $-4.51\pm0.16$& 7&$-3.96\pm0.09$&22\\ 
  21.4 &$-4.16\pm0.05$ & 57&$-4.81\pm0.13$ &11&$-4.64\pm0.10$&17&$-4.34\pm0.07$& 39 \\ 
  21.8 &$-4.30\pm0.04$ & 131& $-4.71\pm0.06$ &49&$-4.60\pm0.05$&65&$-4.59\pm0.05$&66\\ 
  22.2 &$-4.67\pm0.04$ &132&$-4.91\pm0.05$ &81&$-4.81\pm0.04$& 98&$-5.32\pm0.08$&28 \\
  22.6 &$-4.98\pm0.04$ &102&$-5.12\pm0.05$ &74&$-5.06\pm0.05$ &84&$-5.98\pm0.12$&  12 \\
  23.0 &$-5.25\pm0.06$ & 53&$-5.41\pm0.07$ & 38&$-5.33\pm0.06$ & 44&$-6.48\pm0.18$&6\\ 
  23.4 &$-5.73\pm0.09$ & 21& $-5.76\pm0.10$ &19&$-5.73\pm0.09$ & 21& &\\
  23.8 &$-5.97\pm0.13$ & 11&  $-6.36\pm0.17$&  6&$-6.24\pm0.16$&7 & $-6.67\pm0.31$ &2     \\  
  24.2 &$-6.51\pm0.21$ & 4&  $-6.51\pm0.21$ & 4 &$-6.51\pm0.21$ & 4& &&     \\ 
\noalign{\vglue 0.2em}
\noalign{\hrule}
\end{tabular}
\end{center}
\end{table*}

With the aim of investigating this result in more detail, we have then
divided radio sources according to their spectral classification, and
evaluated the local LF for each population taken individually. As in the case for the 
whole sample, values for the different classes of objects are illustrated in Table~1. 
Fig.\ref{fig:lum_fcntype} shows the results for the early-type
(lower panel), early and E+AGN (middle panel) and late-type+SB (top
panel) galaxies. The dashed lines in the bottom and middle panels of
the Fig. are given by equation (\ref{eq:phiell}) which only
considers the local density of `classical', AGN-fuelled,
steep-spectrum radio sources. In this case the agreement with the data
is very good down to powers $P\simeq 10^{20.5} \whzsr$, 
especially if one includes all the objects which show
spectra typical of early-type galaxies, regardless the presence of AGN
emission lines. 

Late-type/SB galaxies do not show the same power-law trend for the LF
as the `classical' radio sources (top panel of Fig.
\ref{fig:lum_fcntype}). Their LF shows a break at about 
$P\simeq 10^{22} \whzsr$, beyond which the
contribution of this class of objects becomes rapidly negligible. For
luminosities fainter than $P\simeq 10^{22}$ however we note that
their spatial density is comparable to that of early/E+AGN-type
galaxies. It follows that, for $P\simeq 10^{21.5}$, late-type galaxies
and starbursts make for about half of the whole spectroscopic sample.

Note that the sharp break in luminosity explains why late-type
radio-emitting sources are only found in the nearby universe, as
already noticed in Section 3. In fact, the majority of these objects
are not bright enough to show radio fluxes $S\ge 1$~mJy beyond
redshifts $z\simeq 0.1$ and therefore were not observed by the FIRST
survey.  This feature in the LF also provides an explanation for the
flattening -- obtained for $b_J\simgt 17.5$ -- of the curves describing
the number of sources (both in the case of the photometric and
spectroscopic samples) per magnitude interval presented in Fig.
\ref{fig:histmags}. As already shown by their radio-to-optical ratios,
late-type/SB galaxies are in fact faint in radio and, being actively
forming stars, bright in the optical b$_J$-band.  The absence of this
class of sources for $P\simgt 10^{22} \whzsr$ as
shown by their LF, implies that they fall out of the spectroscopic
(and photometric) sample for $z\simgt 0.1$ {\it only} because of their
faint radio fluxes and not due to optical magnitudes fainter than the
chosen threshold. These objects would obviously appear as optically
faint at higher redshifts, but they never reach this stage since they
first disappear from the flux-limited radio survey.  It therefore
follows that, for $b_J\simgt 17.5$, only members of the population of
early-type and E+AGN galaxies are included in both the photometric and
spectroscopic samples, which explains the flattening of Fig.
\ref{fig:histmags}.     

Finally, note that the last data point at the bright end of the LF for 
late-type and SB galaxies rises up towards higher spatial densities than 
predicted. Closer investigation reveals that the two objects responsible 
for this measurement, even though classified as late-type galaxies on the 
basis of their line-emission ratios, show spectra where the relative 
importance of the [O${\rm II}$]-[O${\rm III}$] lines seems to point out 
to an `intermediate' case where both AGN and star-formation activity are 
present within the same galaxy. This could then lead to a radio signal 
dominated by non-thermal emission, bringing these sources in the 
class of E+AGN galaxies. 

When it comes to a comparison between observed and predicted LF for the
population of late-type galaxies and starbursts, one has that 
predictions from Dunlop \& Peacock (1990) (not shown for the 
reasons explained earlier in this Section) only agree with the 
$P\simeq 10^{23} \whzsr$ data point, grossly overestimating 
the contribution of this kind of sources at fainter luminosities. 
A better job is provided by the Rowan-Robinson et al. (1993) model, which can 
correctly reproduce both the broken power-law behaviour and the break 
luminosity. However, the slope of the faint-end portion of the predicted LF 
is too steep and, as previously noticed, this results in an overprediction of 
the number density of sources at these luminosities.

By using equation (\ref{eq:phisp}), we find that the best fit is obtained for 
$P_*=10^{21.7} \whzsr$, $\beta=1.4$, $C=0.425\times 
10^{-4}$ and $\sigma=0.44$ (long/short-dashed lines in Fig. 
\ref{fig:lum_fcntype}), not far from the Rowan-Robinson et al. (1993) model, 
showing that late-type/SB galaxies have to be identified with those spirals 
and starbursts which dominate the counts at 60~$\mu$m. 
The above functional form can also, together with the LF 
given by Dunlop \& Peacock (1990) for the population of ellipticals 
and represented by equation (\ref{eq:phiell}), correctly 
reproduce the total LF for radio sources in the 2dFGRS as illustrated by the 
long/short-dashed line in Fig. \ref{fig:L_z}. The agreement however breaks 
down for $P\simlt 10^{20.5} \whzsr$, due to the 
low space density of early- and E+AGN-type galaxies at such faint luminosities.

\section{Redshift distribution}

\begin{figure*}
\vspace{16cm}  
\includegraphics{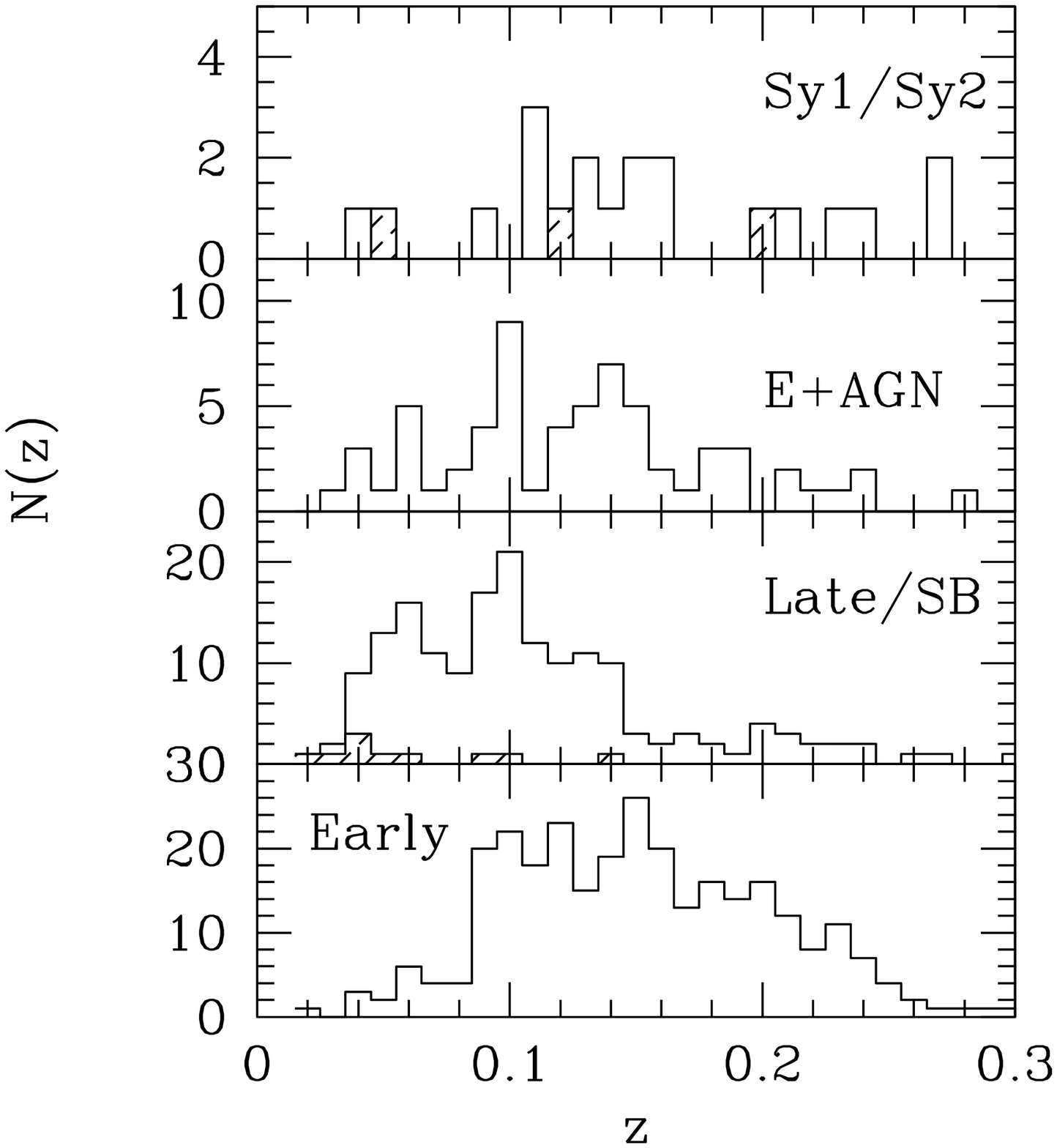}
\includegraphics{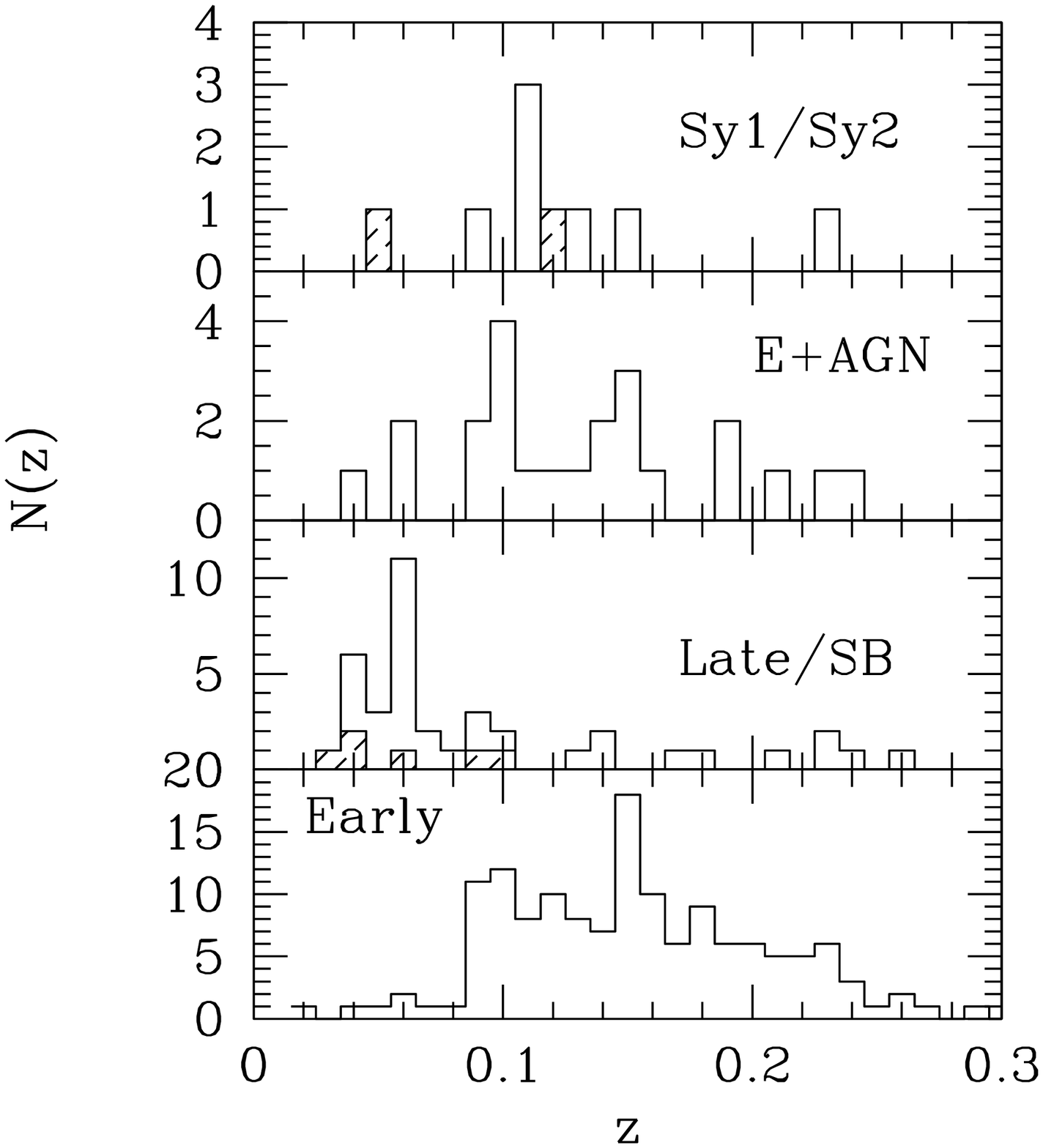}
\includegraphics{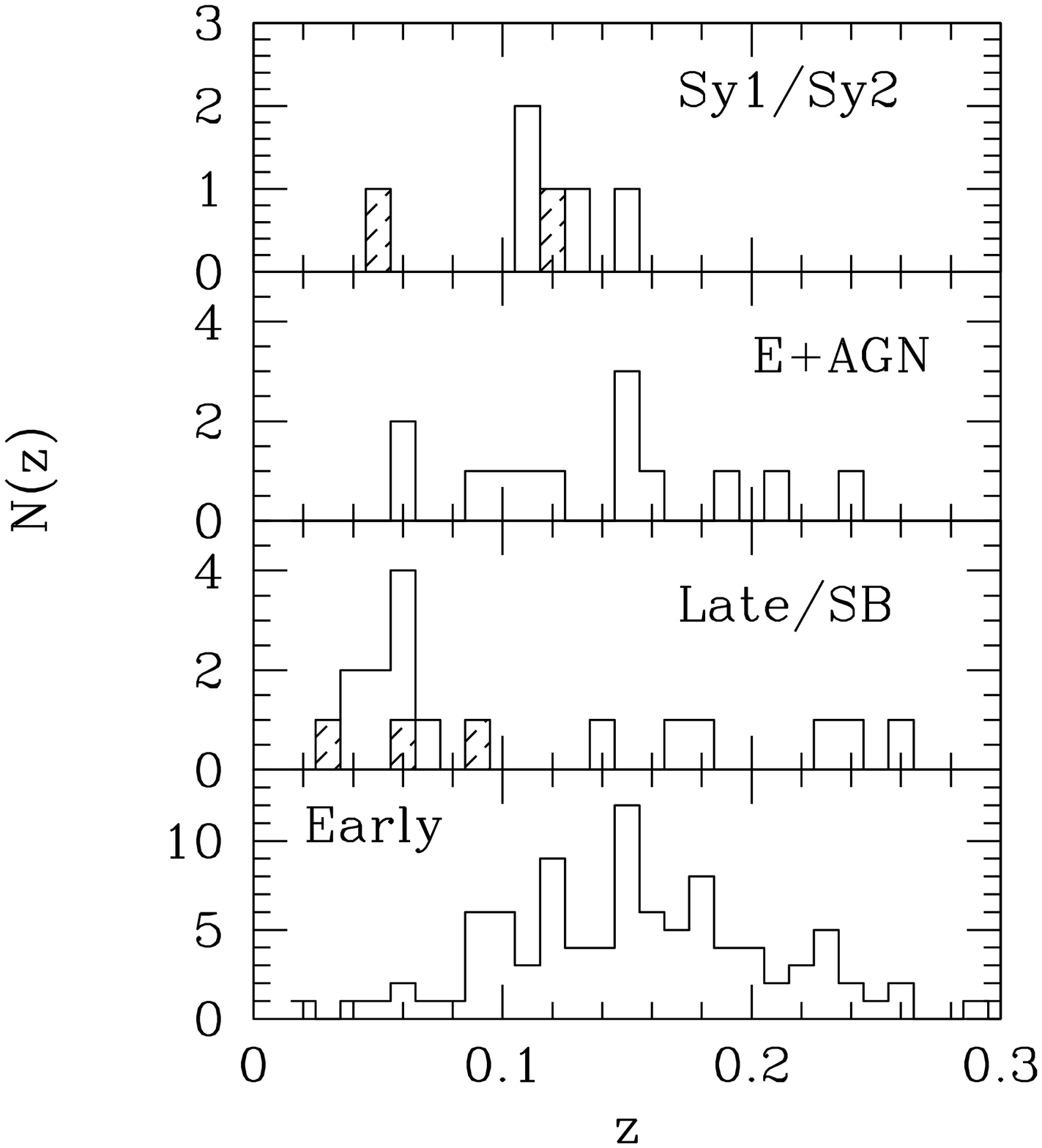}
\includegraphics{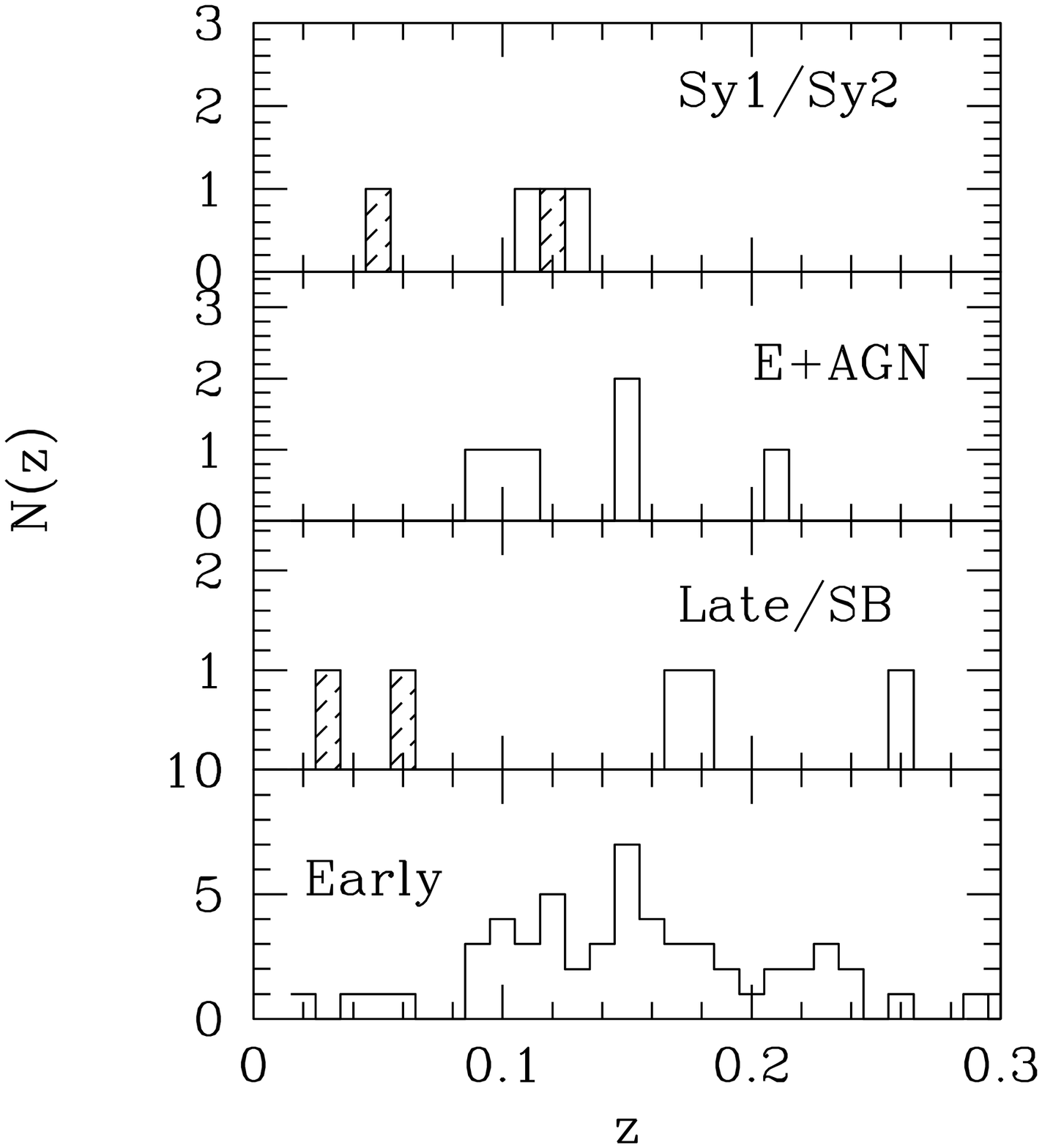}
\caption{Redshift distribution for the different classes of sources in the 
spectroscopic sample. Top left for $S\ge 1$~mJy, top right for $S\ge 3$~mJy,
bottom left $S\ge 5$~mJy and bottom right for $S\ge 10$~mJy. In the Sy1/Sy2 
panels, the shaded histograms represent the distribution of Seyfert 1 
galaxies, while in the Late/SB panels, the shaded histograms are for the 
sub-class of starburst galaxies. 
\label{fig:N_z_type}}   
\end{figure*}

\begin{figure*}
\vspace{14cm}  
\includegraphics{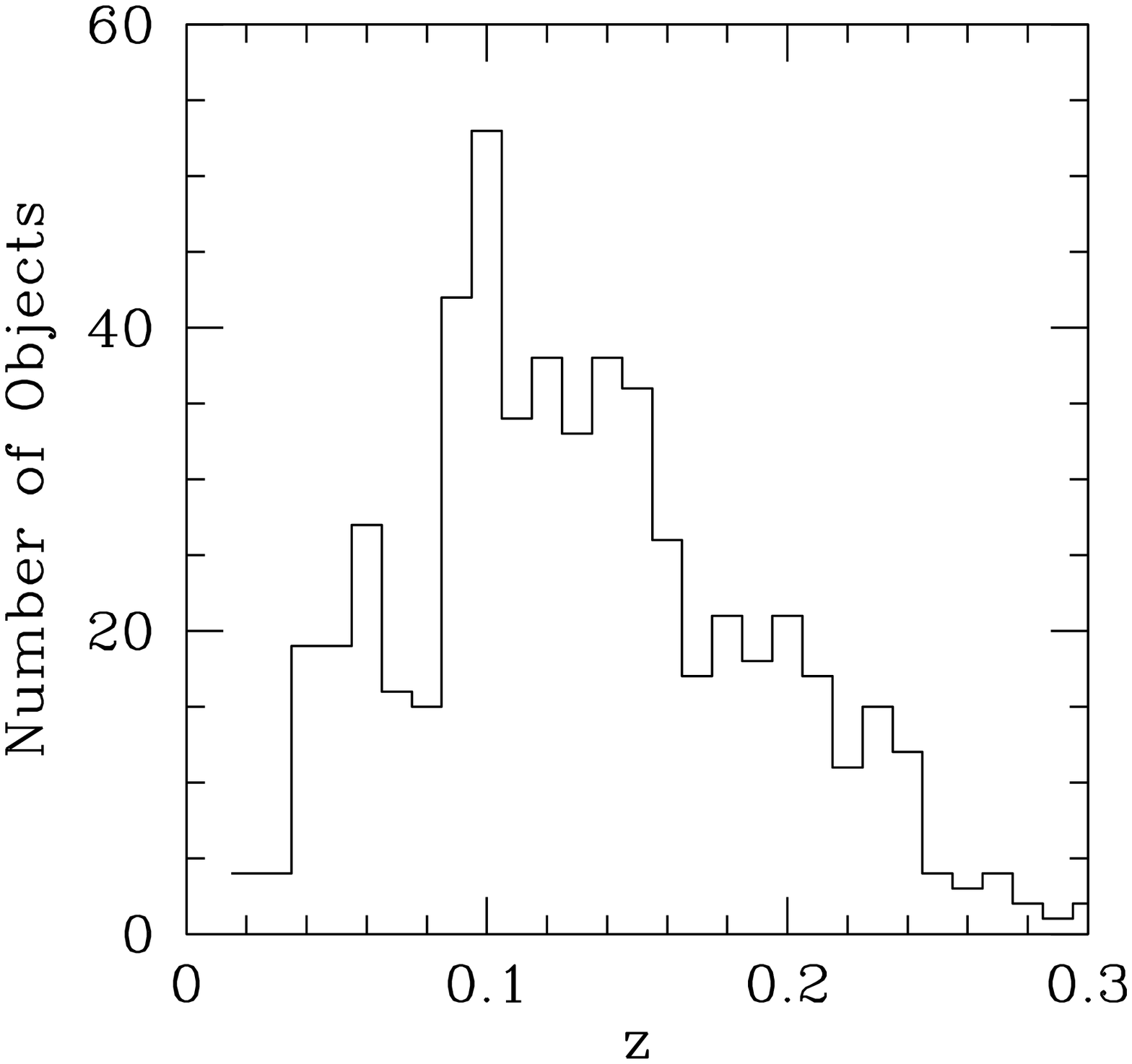}
\includegraphics{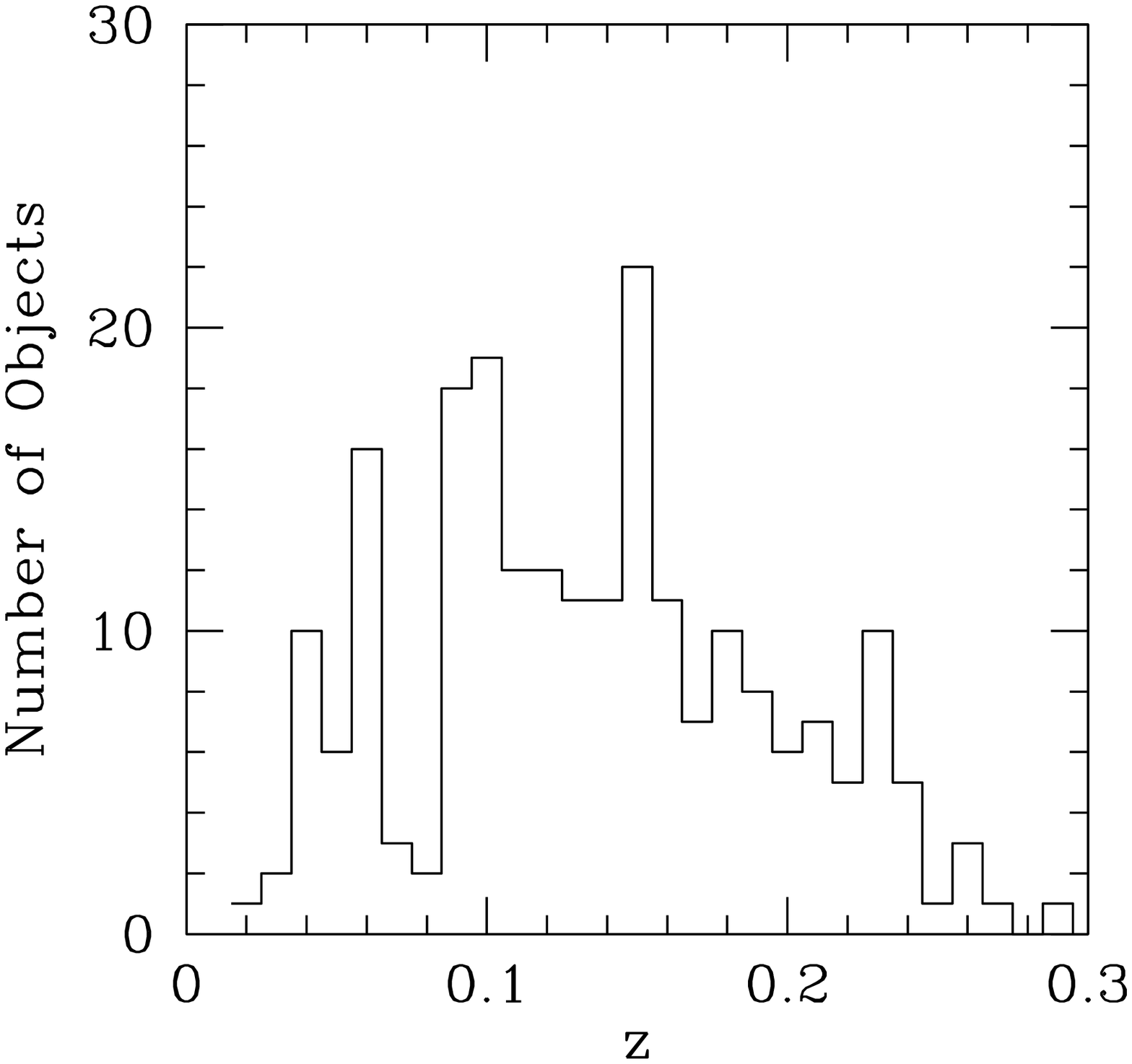}
\includegraphics{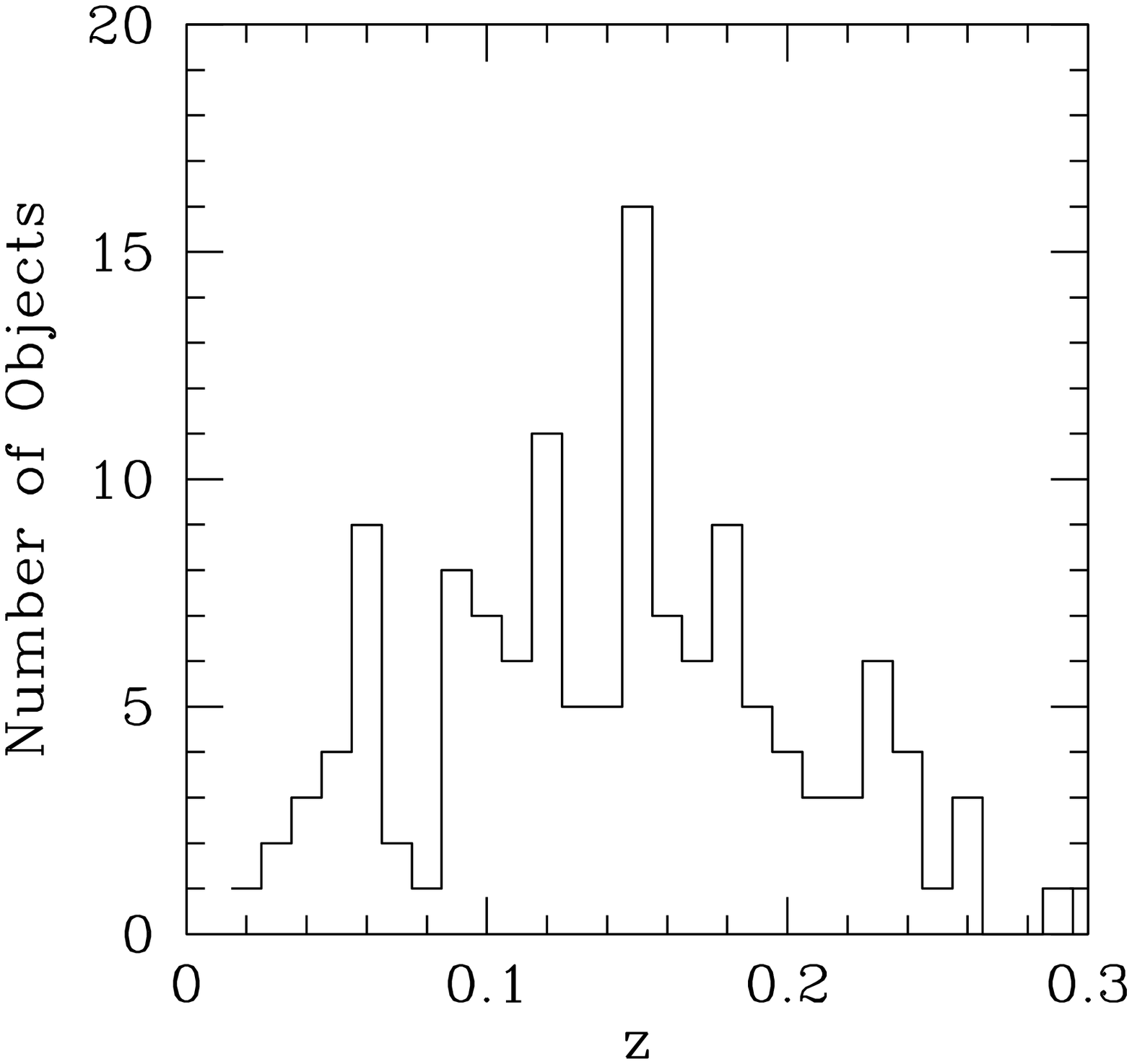}
\includegraphics{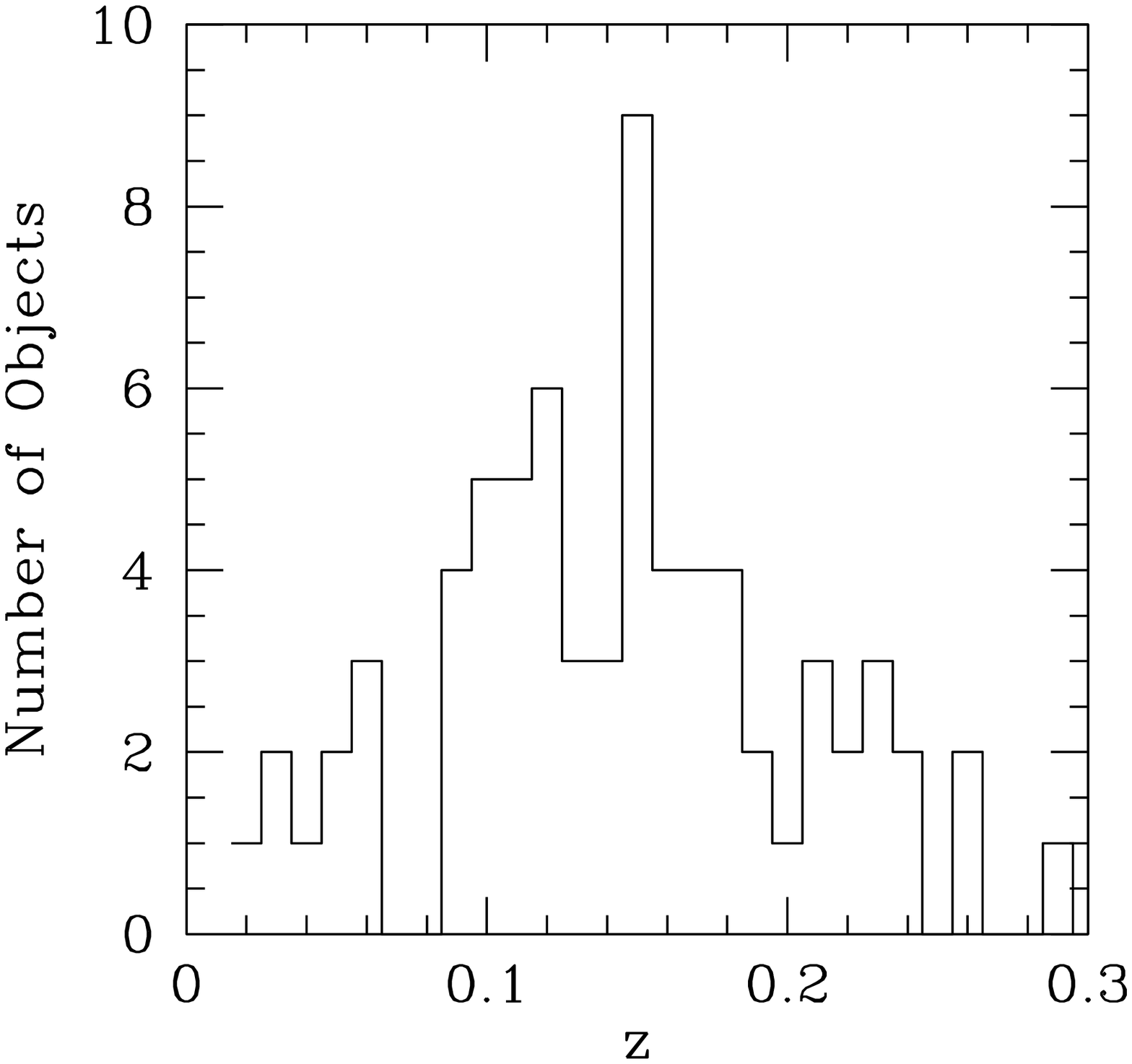}
\caption{Redshift distribution for the whole spectroscopic sample up to 
$z=0.3$. Top left for $S\ge 1$~mJy, top right for $S\ge 3$~mJy,
bottom left $S\ge 5$~mJy and bottom right for $S\ge 10$~mJy.
\label{fig:N_z}}   
\end{figure*}

Finally, we move on to the analysis of the redshift distribution of
sources in the spectroscopic sample. Fig. \ref{fig:N_z_type} shows
the results for different classes of objects and different flux
cuts. In the top-left panel we show the $S\ge 1$~mJy case, in the
top-right panel the $S\ge 3$~mJy case, while the bottom-left panel
only includes those objects with $S\ge 5$~mJy and the bottom-right
panel sources with $S\ge 10$~mJy.

The most striking feature in the various distributions is the drastic
reduction of late-type galaxies for radio fluxes between 1 and 3~mJy.
The tendency for these sources to be only present at $\sim 1$~mJy
level and fainter was already reported by e.g. Windhorst et al. (1985),
Danese et al. (1987), Benn et al. (1993), even though their relative
contribution to the total radio population was still an issue under
debate. As already noted, one also sees that the distribution of
late-type galaxies is very local, peaking at about $z\simeq 0.1$ and
quickly fading off beyond redshifts $z\simeq 0.15$.

Early-type galaxies are instead more distant, only starting to be the
dominant population for $z\simgt 0.1$. They also show a distribution
which does not alter its shape for increasing limiting fluxes. E+AGN
sources once again constitute an intermediate case between late- and
early-type galaxies, and their number is drastically reduced only for
fluxes $S\ge 5$~mJy.

As expected, Seyfert 2 galaxies all present low radio fluxes and are
evenly distributed between $z=0$ and $z=0.3$.  Of the three Seyfert 1
included in the spectroscopic sample within $z\le 0.3$, only one has
$S\le 3$~mJy.

The combination of these trends for the different classes of objects
gives rise to the {\it total} redshift distribution of $S\ge 1$~mJy
radio sources within $z\le 0.3$, although we note that the sample
looses completeness beyond $z\simeq0.2$. This distribution is
presented in Fig. \ref{fig:N_z}; as in Fig.
\ref{fig:N_z_type}, the flux limit in the panels
increases clockwise from the top-left panel between $S=1$ and
$S=10$~mJy. The $z\simeq 0.05$ `bump' shown by all the $N(z)-z$
histograms is principally due to late-type galaxies (with some
contamination from E+AGN), while the maximum at $z\simeq 0.1$ which is
only present in the $S\ge 1$~mJy plot, is due to superposition of
late-type and early-type galaxies in that redshift range. For $S\ge
3$~mJy we see a rather sharp bimodal distribution, with a first
maximum -- principally due to very local star-forming galaxies which
rapidly looses importance as one goes to higher flux cuts -- and by a
smooth increase in the number of sources with redshift between $z=0.1$
and $z=0.2$, -- mainly classical AGN-fuelled galaxies. Above $z=0.2$
the optical apparent magnitude limit means the sample looses the
fainter galaxies.

Fig.~\ref{fig:N_z} shows that the contribution of actively
star-forming galaxies (spirals and irregulars) to the local radio
population is less important than previously thought (see
e.g. Dunlop \& Peacock 1990 where the distributions in redshift
derived from their seven models for sources with $S\ge 1$~mJy show
remarkable `spikes' for  $z \simlt 0.05$ as shown by Magliocchetti et
al. 2000). Our findings however agree with the results from 
Jackson \& Wall 1999 who claim $\sim 30$ per cent of
the sources at 1~mJy to be starburst/starforming galaxies. As the analysis
performed in Sections 3 and 4 indicates, these sources are then most 
likely to be identified
with the radio counterparts of the dusty/starforming IRAS galaxies.

Unfortunately, the question of whether starforming objects are strongly
evolving with redshift (as for instance claimed by Rowan-Robinson et
al. 1993) or not (see e.g. Dunlop \& Peacock 1990) has to be left
unanswered since the limited redshift range of the spectroscopic
sample does not allow any evolutionary analysis. We note however that
the fact that late-type galaxies in the spectroscopic sample are in
general found quite locally, seems to argue against positive
luminosity evolution, which would make these sources brighter at higher
redshifts and therefore observable by a $S\ge 1$~mJy-limited
radio survey such as FIRST beyond $z\sim 0.1$.

\section{Conclusions}
In this work we have used data from the 2dF Galaxy Redshift Survey to 
derive the properties of local $S\ge 1$ mJy radio sources. 
The radio sample was drawn from the FIRST catalogue and all the objects lie 
within the area $9^h 48^m \simlt {\rm RA}({\rm 2000}) 
\simlt 14^h 32^m$ and $-2.77^\circ \simlt {\rm dec}({\rm 2000}) \simlt 2.25^\circ$, 
where the FIRST and APM surveys overlap.
971 radio sources were found to have optical counterparts 
brighter than $b_J=19.45$ in the APM Galaxy catalogue; the 2dFGRS has then 
provided spectra and redshift measurements for 557 of them (the spectroscopic 
sample), up to $z\simeq 0.3$. We note that this apparent incompleteness is 
merely due to incomplete sky coverage of the spectroscopic survey, since 
neither radio nor magnitude biases have been found in the determination of 
both the optical and spectroscopic counterparts of FIRST radio sources. 

Sources in the spectroscopic sample have been divided into three broad
classes on the basis of their spectral appearance:
\begin{enumerate}
\item ``Classical'' radio galaxies (i.e. sources with a spectrum typical of 
absorption systems plus the eventual presence of emission lines due to AGN 
activity); 350 objects corresponding to 63 per cent of the sample, appearing 
for $z\simgt 0.1$ and showing relative high radio-to-optical ratios, red 
colours and radio powers
$10^{21}\simlt P_{1.4{\rm GHz}} / \whzsr \simlt 10^{24}$. 
These objects -- likely to be classical FRI sources -- also present absolute magnitudes 
distributed within a narrow 
interval around $M_B\simeq -21.3$, result which highlights their 
`standard-candle' nature.
\item Late-type galaxies and starbursts; 177 objects, corresponding to about 
32 per 
cent of the spectroscopic sample, mainly present in the very nearby universe 
($z\simlt 0.1$), showing blue colours, faint radio luminosities 
($P_{1.4{\rm GHz}}
\simlt 10^{21.5} \whzsr$) and low radio-to-optical ratios ($r\simlt 10^3$).
\item Seyfert galaxies; 18 Seyfert 2 and 4 Seyfert 1 
(one at $z=2.24$) are included in the spectroscopic sample.
\end{enumerate}

When available, radio and optical images have also allowed
morphological studies of the sources in the sample. As an interesting
result, we find that the majority of the radio sources observed by the
2dFGRS as merging/interacting systems, present spectra typical of
early-type galaxies suggesting that, under appropriate conditions,
galaxy-galaxy interaction can trigger AGN activity even at low
redshifts.
 
Analysis of the local radio luminosity function for the spectroscopic sample 
shows that, for $P_{1.4{\rm GHz}}\simgt 10^{20.5} \whzsr$, 
classical radio galaxies are well described by models such as the one introduced by
Dunlop \& Peacock (1990) for pure luminosity evolution, although we note that 
no estimate on the evolutionary degree of this population was possible due to 
the limited redshift range spanned by the 2dFGRS.

Late-type galaxies instead exhibit a broken power-law trend for their 
luminosity function. Such a break and the rapid decrease of the number density 
of sources for $P_{1.4{\rm GHz}}\simgt 10^{22} \whzsr$ 
explains the quick disappearance of this class of objects beyond $z\simeq 0.1$ as shown by 
their redshift distribution. The observed LF in this case is in agreement with 
the one derived by Saunders et al. (1990) for IRAS galaxies.
This supports the assumption of Rowan-Robinson et al. (1993) that late-type 
radio galaxies should be identified with the population of dusty spirals 
and starbursts observed at 60~$\mu$m. 

\include{important_table1}
\include{important_table2}


\begin{thebibliography}{} 
\bibitem[\protect\citename{Becker et al. }1995]{Be}
Becker R.H., White R.L., Helfand D.J., 1995, {ApJ}, {450}, 559 
\bibitem[\protect\citename{Benn et al. }1993]{Ben}
Benn C.R., Rowan-Robinson M., McMahon R.G., Broadhurst T.J., Lawrence
A., 1993; {MNRAS}, {263}, 98 
\bibitem[\protect\citename{Bland }1978]{Blan}
Blandford R., Rees M.J., 1978, { MNRAS}, { 169}, 395 
\bibitem[\protect\citename{Bock et al. }1999]{Bo}
Bock D. C-J., Large M.I., Sadler, E.M., 1999, AJ, 117, 1578
\bibitem[\protect\citename{Colless }2001]{Col}
Colless M. et al. (2dFGRS team), 2001, MNRAS, 328, 1039
\bibitem[\protect\citename{Condon }1984]{Con}
Condon J.J., 1984, { ApJ}, { 287}, 461 
\bibitem[\protect\citename{Condon et al. }1998]{Co}
Condon J.J., Cotton W.D., Greisen E.W., Yin Q.F., Perley R.A., Taylor G.B.,
Broderick J.J., 1998, AJ, 115, 1693 
\bibitem[\protect\citename{Danese }1987]{Da}
Danese L., De Zotti G., Franceschini A., Toffolatti L., 1987; {
  ApJ}, {318}, L15   
\bibitem[\protect\citename{Dunlop }1990]{Dun}
Dunlop J.S., Peacock J.A., 1990,{ MNRAS}, {247}, 19  
 \bibitem[\protect\citename{FR }1974]{FR}
Fanaroff B.L., Riley J.M., 1974, { MNRAS}, { 167}, L31       
\bibitem[\protect\citename{Folkes et al. }1999]{Folk}
Folkes S.R., Ronen S., Price I., Lahav O., Colless M., Maddox S.J., 
Glazebrook K., Bland-Hawthorn J, Cannon R., Cole S., Collins C., Couch W., 
Driver S., Dalton G., Efstathiou G., Ellis R., Frenk C., Kaiser N., Lewis L., 
Lumsden S., Peacock J., Peterson B., Sutherland W., Taylor K., 1999, MNRAS, 
308, 459.
\bibitem[\protect\citename{Georgakakis et al. }1999]{Geo}
Georgakakis A., Mobasher B., Cram L., Hopkins A., Lidman C,
Rowan-Robinson M., 1999; {MNRAS}, {306}, 708
\bibitem[\protect\citename{Glaz et al. }1998]{Glaz}
Glazebrook K., Offer A.R., Deeley K., 1998, ApJ, 492, 98
\bibitem[\protect\citename{Gruppioni et al. }1998]{grup}
Gruppioni C., Mignoli M., Zamorani G., 1999; {MNRAS}, {304}, 199
\bibitem[\protect\citename{Hill }2001]{Hill}
Hill T.L., Heisler C.A., Norris R.P., Reynolds J.E., Hunstead R.W., 2001, 
ApJ, 548, 127  
\bibitem[\protect\citename{wa et al. }1999]{wa}
Jackson C.A., Wall J.V., 1999, { MNRAS}, { 304}, 160
\bibitem[\protect\citename{Jackson et al. }2001]{Jack}
Jackson, C.A., 2002, in preparation
\bibitem[\protect\citename{Kennicut }1992]{Kennicut}
Kennicut R.C., 1992, ApJ, {388}, 310   
\bibitem[\protect\citename{Kochanek }2001]{Kochanek}
Kochanek C.S., Pahre M.A., Falco E.E., 2001, ApJ, submitted, astro-ph/0011458 
\bibitem[\protect\citename{Led }1996]{Led}
Ledlow M.J., Owen F.N., 1996, AJ, 112, 9
\bibitem[\protect\citename{Maddox et al. }1990a]{Mada}
Maddox S.J., Efstathiou G., Sutherland W.J., Loveday J., 1990a, {
  MNRAS}, {243}, 692        
\bibitem[\protect\citename{Maddox et al. }1990b]{Madb}
Maddox S.J., Efstathiou G., Sutherland W.J., 1990b, {
  MNRAS}, {246}, 433    
\bibitem[\protect\citename{Maddox et al. }1996]{Madc}
Maddox S.J., Efstathiou G., Sutherland W.J., 1996, {
  MNRAS}, {283}, 1227
\bibitem[\protect\citename{Maddox et al. }1998]{Maddox} 
Maddox S.J., 1998, in `Large-Scale Structure: Tracks and Traces', 
proc. 12th Postdam Cosmogony Workshop, World Scientific, 91, astro-ph/9711015
\bibitem[\protect\citename{Madwick et al. }2001]{Madwick} 
Madgwick D. et al. (2dFGRS team), 2001, MNRAS, submitted, astro-ph/0107197
\bibitem[\protect\citename{Magliocchetti et al. }1998]{Mag}
Magliocchetti M., Maddox S.J., Lahav O., Wall J.V., 1998, {MNRAS}, {
  300}, 257
\bibitem[\protect\citename{Maglioa }2001]{Maglioa}
Magliocchetti M., Maddox S., Wall J.V., Benn C.R., Cotter G., 2000, MNRAS, 
318, 1047 
\bibitem[\protect\citename{Maglio }2001]{Maglio} 
Magliocchetti M., Maddox S.J., 2002, MNRAS, 330, 241
\bibitem[\protect\citename{Maraschi }1994]{mara} Maraschi L., Rovetti F., 
1994, { ApJ}, { 436}, 79   
\bibitem[\protect\citename{Masci et al. }2001]{Masci}
Masci F.J., Condon J.J., Barlow T.A., Lonsdale C.J., Xu C., Shupe D.L., Pevunova O., Fang F., 
Cutri R., 2001, PASP, 113,10
\bibitem[\protect\citename{Mc }1995]{Mc}
McQuade K., Calzetti D., Kinney A.L., 1995, ApJS, 97, 331
\bibitem[\protect\citename{Orr }1982]{orr}
Orr M.J.L., Browne I.W.A., 1982, { MNRAS}, { 200}, 1067   
\bibitem[\protect\citename{Ort }1987]{ort}
Oort M.J.A., Steemers W.J.G., Windhorst R.A., 1987, { A\&AS}, { 73}, 103
\bibitem[\protect\citename{Padovani }1992]{padu}
Padovani P., Urry, C.M., 1992, { ApJ}, { 387}, 449        
\bibitem[\protect\citename{Rixon et al. }1997]{Rix}
Rixon G.T., Wall J.V., Benn C.R., 1991; {MNRAS}, {251},
243 
\bibitem[\protect\citename{Rola }1997]{Rola}
Rola C., Terlevich E., Terlevich R., 1997, MNRAS, 289, 419 
 \bibitem[\protect\citename{Rowan1 et al. }1968]{Rowan}
Rowan-Robinson M., 1968, MNRAS, 138, 445
 \bibitem[\protect\citename{Rowan et al. }1993]{Row}
Rowan-Robinson M., Benn C. R., Lawrence A., McMahon R.G., Broadhurst
T.J., 1993, { MNRAS}, { 23}, 123      
\bibitem[\protect\citename{Sadler et al. }1999]{sa}
Sadler E.M., McIntyre V.J., Jackson, C. A., Cannon R.D., 1999, PASA, 16, 247 
\bibitem[\protect\citename{Sadler et al. }2002]{sadl}
Sadler E.M. et al. (2dFGRS team), 2002, MNRAS, 329, 227
\bibitem[\protect\citename{Sau et al. }1990]{sau}
Saunders W., Rowan-Robinson M., Lawrence A., Efstathiou G., Kaiser N., Ellis 
R.S., Frenck C.S., 1990, MNRAS, 242, 318 
\bibitem[\protect\citename{Veill }1987]{Veill}
Veilleux S., Osterbrock D.E., 1987, ApJ, 63, 295
\bibitem[\protect\citename{wall et al. }1980]{wall}
Wall J.V., Pearson T.J., Longair M.S., 1980, { MNRAS}, { 193}, 686       
\bibitem[\protect\citename{wa et al. }1997]{wall2}
Wall J.V., Jackson C.A., 1997, { MNRAS}, { 290}, L17     
\bibitem[\protect\citename{Wolt }1990]{Wolt}
Woltjer L., 1990, in SAAS-FEE Advanced Course on Active Galatic Nuclei, 
Swiss Society for Astro-Physics Astronomy, eds. Blandford R., Netzer H., 
Woltjer L.   
\bibitem[\protect\citename{Windhorst et al.}1985]{Win}
Windhorst R.A., Miley G.K., Owen F.N., Kron R.G., Koo R.C., 1985;
{ApJ}, {289}, 494   
\end{thebibliography}
\end{document}